\documentclass[useAMS,usegraphicx,usenatbib]{mn2e}
\usepackage{lscape}

\title[Bayesian search for low-mass planets around nearby M dwarfs]{Bayesian search for low-mass planets around nearby M dwarfs -- Estimates for occurrence rate based on global detectability statistics}

\author[M. Tuomi et al. 2014]{Mikko Tuomi$^{1,2}$\thanks{E-mail: \texttt{mikko.tuomi@utu.fi}; \texttt{m.tuomi@herts.ac.uk}}, Hugh R. A. Jones$^{1}$, John R. Barnes$^{1}$, Guillem Anglada-Escud\'e$^{3}$, \and and James S. Jenkins$^{4}$\\
\\
$^{1}$University of Hertfordshire, Centre for Astrophysics Research, Science and Technology Research Institute, College Lane, AL10 9AB,\\ Hatfield, UK\\
$^{2}$University of Turku, Tuorla Observatory, Department of Physics and Astronomy, V\"ais\"al\"antie 20, FI-21500, Piikki\"o, Finland\\
$^{3}$Queen Mary, University of London, Astronomy Unit, School of Mathematical Sciences, London, UK\\
$^{4}$Departamento de Astronom\'ia, Universidad de Chile, Camino del Observatorio 1515, Las Condes, Santiago, Chile}

\begin{document}

\date{Accepted {XX.XX.2014}. Received {XX.XX.2014}; in original form {XX.XX.2014}}

\pagerange{\pageref{firstpage} -- \pageref{lastpage}} \pubyear{2014}

\maketitle

\label{firstpage}

\begin{abstract}
Due to their higher planet-star mass-ratios, M dwarfs are the easiest targets for detection of low-mass planets orbiting nearby stars using Doppler spectroscopy. Furthermore, because of their low masses and luminosities, Doppler measurements enable the detection of low-mass planets in their habitable zones that correspond to closer orbits than for Solar-type stars. We re-analyse literature UVES radial velocities of 41 nearby M dwarfs in a combination with new velocities obtained from publicly available spectra from the HARPS-ESO spectrograph of these stars in an attempt to constrain any low-amplitude Keplerian signals. We apply Bayesian signal detection criteria, together with posterior sampling techniques, in combination with noise models that take into account correlations in the data and obtain estimates for the number of planet candidates in the sample. More generally, we use the estimated detection probability function to calculate the occurrence rate of low-mass planets around nearby M dwarfs. We report eight new planet candidates in the sample (orbiting GJ 27.1, GJ 160.2, GJ 180, GJ 229, GJ 422, and GJ 682), including two new multiplanet systems, and confirm two previously known candidates in the GJ 433 system based on detections of Keplerian signals in the combined UVES and HARPS radial velocity data that cannot be explained by periodic and/or quasiperiodic phenomena related to stellar activities. Finally, we use the estimated detection probability function to calculate the occurrence rate of low-mass planets around nearby M dwarfs. According to our results, M dwarfs are hosts to an abundance of low-mass planets and the occurrence rate of planets less massive than 10 M$_{\oplus}$ is of the order of one planet per star, possibly even greater. Our results also indicate that planets with masses between 3 and 10 M$_{\oplus}$  are common in the stellar habitable zones of M dwarfs with an estimated occurrence rate of 0.21$^{+0.03}_{-0.05}$ planets per star.
\end{abstract}



\section{Introduction}

In recent years, planets have been discovered around the least massive stars, M dwarfs, in a diversity of different configurations with widely varying orbital properties and masses \citep[e.g.][and references therein]{endl2006,bonfils2013}. For instance, there are several high-multiplicity systems around M dwarfs consisting of only low-mass planets that can be referred to as super-Earths or Neptunes, such as those orbiting GJ 581 \citep{bonfils2005,udry2007,mayor2009}\footnote{We note that the number of planets around GJ 581 is uncertain with different authors reporting different numbers from three to six \citep[see][]{vogt2010,vogt2012,gregory2011,tuomi2011,baluev2012}.}, GJ 667C \citep{anglada2012,anglada2013,delfosse2012}, and GJ 163 \citep{bonfils2013b,tuomi2013c}. Recent precision velocity surveys have also revealed the existence of more massive planetary companions orbiting nearby M dwarfs \citep[e.g.][]{rivera2010,anglada2012b} showing that such companions do exist, but not in abundance \citep{bonfils2013,montet2013}, and are less common than for K, G, and F stars \citep{endl2006}. However, the most interesting planetary companions around these stars are the low-mass ones that orbit their hosts with such separations that, under certain assumptions regarding atmospheric properties, they can be estimated to enable the existence of water in its liquid form on the planetary surfaces \citep[e.g.][]{selsis2007,kopparapu2013}. Planets of this type -- sometimes called habitable-zone super-Earths -- are easier to detect around M dwarfs than around more massive stars because the planet-star mass-ratios give rise to signals with sufficiently high amplitudes, and the shorter orbital periods allow for more orbital phases to be sampled in data covering a fixed length of time, to enable their detections \citep[e.g.][]{mayor2009,anglada2013,tuomi2013c}.

Recently, accurate estimates for the occurrence rate of planets in the \emph{Kepler's} field have been reported in several studies \citep[e.g.][]{howard2012,dressing2013,morton2013}. One of the most interesting features in the \emph{Kepler} sample is that the occurrence rate of planets around stars appears to increase from roughly 0.05 planets per star around F2 stars to 0.3 per star around M0 dwarfs \citep{howard2012}, although the functional form of this relation is far from certain. This increase applies to planets with orbital periods below 50 days because of the available baseline of the \emph{Kepler} data. While \emph{Kepler} will be able to provide occurrence rates for longer orbital periods, possibly up to 200-300 days, radial velocity surveys will be needed to probe the occurrence rate of planets on orbits longer than that. Moreover, unlike planets around more massive K, G, and F stars that have been targeted by the \emph{Kepler} space-telescope in abundance, M dwarfs are not bright enough to be found in comparable numbers in the \emph{Kepler's} field, which makes it difficult to estimate the occurrence rates and statistical properties of planets around such stars in detail.

According to \citet{dressing2013}, the \emph{Kepler's} sample contains 3897 stars with estimated effective temperatures below 4000 K, out of which 64 are planet candidate host stars with a total of 95 candidate planets orbiting them. \citet{dressing2013} concluded that with periods ($P$) less than 50 days, the occurrence rate of planets with radii between 0.5 R$_{\oplus} < r_{p} < 4 $R$_{\oplus}$ is 0.90$^{+0.04}_{-0.03}$ planets per star; with radii between 0.5 R$_{\oplus} < r_{p} <$ 1.4 R$_{\oplus}$ is 0.51$^{+0.13}_{-0.06}$ planets per star, although this estimate might be underestimated as much as by a factor of two \citep{morton2013}; and that the occurrence rate of planets with $r_{p} > 1.4$ R$_{\oplus}$ decreases as a function of decreasing stellar temperature. Furthermore, the occurrence rate of planets appears to decrease heavily between 2 and 4 R$_{\oplus}$, which is indicative of overabundance of planets with low radii and therefore low masses \citep{morton2013}. These findings challenge the results obtained using radial velocity surveys that should be able to detect planets with similar statistics, although the comparison with \emph{Kepler's} results is difficult due to the challenges in comparing populations described in terms of planetary radii and minimum masses in the absense of accurate population models for planetary compositions and therefore densities. The estimates based on transits detected by using the \emph{Kepler} telescope might also be contaminated by a false positive rate of $\sim$ 10\% due to astrophysical effects such as stellar binaries in the background \citep{morton2011,fressin2013}. 

Far fewer planets around M dwarfs are known from radial velocity surveys of such stars \citep[e.g.][who reported nine planet candidates in their sample]{bonfils2013}. However, the ones that are known are among the richest and the most interesting extrasolar planetary systems in terms of numbers of planets, their orbital spacing and dynamical packedness, and their low masses \citep[e.g.][]{mayor2009,rivera2010,anglada2012b,anglada2013,tuomi2013c}. To a certain extent, this lack of known planets around M dwarfs is due to observational biases arising from the fact that early radial velocity surveys did not target low-mass stars because of the difficulties in obtaining sufficiently high signal-to-noise observations due to a lack of photons in the V band to enable high quality radial velocity measurements. Another reason was that --  based on a sample size of unity -- Solar-type stars were considered more promising hosts to planetary systems. This observational bias is also likely caused by the fact that -- in comparison with stars of the spectral classes F, G, and K -- massive giant planets are not as abundant around M dwarfs \citep{bonfils2013}, and the planets that exist, if they indeed do exist, are likely so small that they induce radial velocity signals that have amplitudes comparable to the current high-precision measurement noise levels, which makes their detection difficult at best.

\citet{bonfils2013} reported estimates for the occurrence rates of planetary companions orbiting M dwarfs based on radial velocity measurements obtained by using the \emph{High Accuracy Radial velocity Planet Searcher} (HARPS) spectrograph. According to their results, super-Earths with minimum masses between 1 and 10 M$_{\oplus}$ are abundant around M dwarfs with an occurrence rate of 0.36$^{+0.25}_{-0.10}$ for periods between 1 and 10 days and 0.52$^{+0.50}_{-0.16}$ for periods between 10 and 100 days, respectively. Furhtermore, they reported an estimate for the occurrence rate of super-Earths in the habitable zones (HZs) of M dwarfs of 0.41$^{+0.54}_{-0.13}$ planets per star.

M dwarfs are the most abundant type of stars in the Solar neighbourhood. Therefore, the occurrence rate of planets around these stars will dominate any general estimates of the occurrence rate of planets. For this reason, we re-analyse the radial velocities obtained using the \emph{Ultraviolet and Visual Echelle Spectrograph} (UVES) at VLT-UT2 of a sample of M dwarfs of \citet{zechmeister2009} using posterior sampling techniques in our Bayesian search for planetary signals. We also extract HARPS radial velocities for these stars from the publicly available spectra in the European Southern Observatory (ESO) archive and analyse the combined UVES and HARPS velocities. The methods are presented in Section \ref{sec:statistical_tools} in detail and we show the results based on combined HARPS and UVES data in Section \ref{sec:UVES}. We present the statistics of the new planet candidates we detect and compare the obtained occurrence rates to other planet surveys targeting M dwarfs in Section \ref{sec:planet_statistics}, describe some of the interesting new planetary systems and the evidence in favour of their existence in greater detail in Section \ref{sec:new_systems}, and discuss the results in Section \ref{sec:discussion}.

\section{Statistical methods and benchmark model}\label{sec:statistical_tools}

We analyse the radial velocity data sets by using posterior sampling algorithms and estimations of Bayesian evidences for models with $k = 0, 1, ...$ Keplerian signals. Throughout the analyses we apply a fully Bayesian data analysis framework as discussed and applied in several astronomy papers over the recent years \citep[e.g.][]{ford2005,ford2006,trotta2007,feroz2011,gregory2011a,loredo2012,tuomi2012,tuomi2013a,tuomi2013b}. In particular, we apply the adaptive Metropolis posterior sampling algorithm of \citet{haario2001} that can be applied readily for analyses of radial velocity data \citep[e.g.][]{tuomi2012,tuomi2013a,tuomi2013b}. We estimate the Bayesian evidences in favour of a given number of signals (i.e. in favour of a model $\mathcal{M}_{k}$ that contains $k$ signals) by using the estimate of \citet{newton1994} based on statistical samples drawn from both the prior and the posterior densities, and report parameter estimates using the maximum \emph{a posteriori} (MAP) estimates and 99\% Bayesian credibility intervals (BCS). We note that the acronym BCS stands for Bayesian credibility set, which is represented by an interval in a single dimension when the posterior density does not have multiple significant modes. This set with a probability threshold of $\delta$ is a set $\mathcal{D}_{\delta}$ defined for a posterior density $\pi(\theta | m)$ as
\begin{eqnarray}
  \mathcal{D}_{\delta} = & \bigg\{ \theta \in C \subset \Omega : \int_{\theta \in C} \pi(\theta | m) d \theta = \delta, \nonumber\\
  & \pi(\theta | m) \big|_{\theta \in \partial C} = c \bigg\},
\end{eqnarray}
where $\Omega$ is the parameter space of the parameter vector $\theta$, $\pi(\theta | m)$ is a posterior probability density function given measurements $m$, $\partial \mathcal{D}_{\delta}$ represents the edge of the set $\mathcal{D}_{\delta}$, and $c$ is some positive constant. Formal definition and discussion can be found in textbooks of Bayesian statistics, e.g. \citet{berger1980} and \citet{kaipio2005}.

Our benchmark statistical model contains linear acceleration and correlation components. In the context of this model, we describe the radial velocity measurement ($m_{i,l}$) made at epoch $t_{i}$ with instrument $l$ as in \citet{tuomi2013a,tuomi2013b} and write it as
\begin{equation}\label{eq:model}
  m_{i,l} = f_{k}(t_{i}) + \gamma_{l} + \dot{\gamma}t_{i} + \epsilon_{i,l} + \phi_{l} \exp \big\{ \alpha (t_{i-1}-t_{i}) \big\} \epsilon_{i-1,l},
\end{equation}
where $\gamma_{l}$ and $\dot{\gamma}$ are free parameters of the model representing the reference velocity of the $l$th instrument and the linear acceleration, respectively, and $\epsilon_{i,j}$ is a Gaussian random variable with a zero mean and variance $\sigma_{i}^{2} + \sigma_{l}^{2}$ describing the amount of Gaussian white noise in the $i$th measurement of the instrument $l$. Parameter(s) $\phi_{l}$ represents the correlation between the deviations of the $i$th and $i-1$th measurement from the mean, i.e. the dependence of the $i$th measurement on the deviation of the $i-1$th measurement from the mean because there can be no causal relationship the other way around\footnote{It is not necessary to assume causality as the model simply aims at removing correlations and is therefore only a statistical model that describes the data reasonably accurately.}. The parameter vector $\theta$ of a ``baseline'' model without Keplerian signals is then $\theta = (\gamma_{l}, \dot{\gamma}, \sigma_{l}, \phi_{l})$ for one instument. In our notation, function $f_{k}$ represents the superposition of $k$ Keplerian signals.

In Eq. (\ref{eq:model}), parameter $\alpha$ corresponds to the timescale of the exponential smoothing in the moving average (MA) component \citep{tuomi2013b}. We chose this parameter such that correlations on the time-scale of few days were taken into account because correlations in this time-scale are known to occur in radial velocities \citep{baluev2012,tuomi2013b} but that measurements sufficiently far in time from one another, i.e. more than a dozen days or so, are unlikely to be correlated. Therefore, we chose $\alpha = 0.01$ hours$^{-1}$ -- a value that was supported by the largest UVES datasets (GJ 551 and GJ 699). This value was also found to describe the data sets well because changing the value resulted in, at most, equally good performance of the statistical model for the data sets with large number of measurements. It was not necessary to treat this parameter as a free parameter of the statistical model because in most of the cases, it could not be constrained at all due to a small number of data points and in such cases the principle of parsimony should be applied to decrease the number of free parameters and e.g. fix the time-scale parameter as we have done in the current work. The value $\alpha = 0.01$ hours$^{-1}$ corresponds to a correlation time-scale of roughly 4 days. It is possible that this choice for the time-scale affects the analysis results of some data sets that in fact have correlations on a much shorter (longer) time-scale of few hours (dozen days) instead of days. However, we consider this to be rather unlikely because of the low number of measurements in most data sets. We also note that there are thus $3j+5k+1$ free parameters in our model if $j$ is the number of instruments that have been used to obtain the data.

While our benchmark model in Eq. (\ref{eq:model}) cannot be expected to be a perfect description of the radial velocities \citep[e.g.][]{tuomi2012c,tuomi2013b}, we still estimate that it contains most of the important features of a good radial velocity model. Particularly, it contains the linear acceleration that could be present due to a previously unknown long-period substellar companion. Even without evidence of such a companion (i.e. of such a trend), we include this linear acceleration in the model to construct a standard model that can be applied to all the data sets we analyse. If there is no linear acceleration in a given data set, its inclusion in the model makes the model overparameterised because dropping the corresponding term from the formulation would provide a better model due to the principle of parsimony. However, we are willing to risk such overparameterisation to be able to use the same general benchmark model for all the data sets and to obtain as trustworthy results as possible. The same principle applies to the parameter $\phi_{l}$ that quantifies the amount of intrinsic correlation in the data obtained using the $l$th instrument \citep[e.g.][]{tuomi2012c,tuomi2013b}. We use this MA component in the model even if we cannot show that it is significantly different from zero, which makes the model more complicated but enables us to see consistently whether the excess noise in the data has a significant red-colour component.

We note that even if there is no evidence for neither acceleration nor correlation in the sense that the parameters $\dot{\gamma}$ and $\phi_{l}$ have posterior densities that are consistent with zero, including the corresponding terms in the model makes the results more robust because the uncertainties of these terms can be taken into account directly. Furthermore, this ensures that none of the signals we detect are spuriously caused by the combination of these two factors.

\subsection{Prior choice}

Because prior densities of the model parameters are an integral part of Bayesian analyses, we discuss our prior choices briefly. Throughout the analyses, we use prior probability densities as described in \citet{tuomi2012} with one small but possibly significant exception. In comparison to the choice of \citet{tuomi2012}, we follow \citet{tuomi2013c} and use more restrictive eccentricity priors in our analyses by adopting $\pi(e) \propto \mathcal{N}(0, 0.1^2)$. In addition to being a more approppriate functional form for an eccentricity prior than a uniform one \citep[see also][for comparison]{kipping2013}, we make this choice because the data sets we analyse have already been analysed by \citet{zechmeister2009} who did not report any planetary signals in the UVES data sets. Therefore, if there are any significant signals in these data sets, they are likely to be deeply embedded in the noise in the sense that their amplitudes are comparable to the noise levels, which results in overestimated orbital eccentricities \citep{zakamska2011}. Yet, because low-mass planets are mostly found on close-circular orbits \citep{tuomi2013c}, we prefer a slight underestimation of orbital eccentricities over their overestimation. We note that this prior choice is still much more conservative than the commonly made decision to fix eccentricities to zero \emph{a priori} that correspond to choosing a delta-function prior such that $\pi(e) = \delta(e)$. For further justification of this prior choice, we refer to Appendix A in \citet{anglada2013}.

There is also the possibility that our eccentricity prior decreases the significance of a signal that is actually caused by e.g. a planet on a moderately eccentric orbit. However, we consider that to be unlikely because in the current sample, any planetary signals we detect have amplitudes that are comparable to the measurement noise. This means that the corresponding eccentricities are ill-constrained from below and that using a prior density as described above, or indeed the choice proposed by \citet{kipping2013}, is thus unikely to affect the results significantly even if some of the stars in the sample have low-mass planets on eccentric orbits.

\subsection{Posterior samplings}

While the adaptive Metropolis algorithm is an efficient tool in drawing statistically representative samples in cases of unimodal posteriors with little non-linear correlations between the parameters, it is not necessarily well-suited for such samplings of multimodal posteriors that are typical, especially with respect to the period parameters of the signals, when searching for periodic signals of planets. Therefore, when searching for a $k$th signal we simply divide the period space of this signal into parts that only contain one significant maximum that would, because of its significance, effectively prevent the chain from ``jumping'' between the different modes. This enables us to use the adaptive Metropolis algorithm. These different parts can then be sampled independently and treated as different (\emph{a priori}) models to assess the relative significances of the corresponding maxima. However, in practice, such cases were rare and we only applied such divisions of the period space to the data of two targets.

When estimating the significances of the solutions, i.e. that the Markov chains were sufficiently close to convergence, we used the Gelman-Rubin statistics \citep{gelman1992} as also described in \citet{ford2006}. In particular, we required that the test statistics $R(\theta)$ that approaches unity from above as the Markov chains approach convergence was below 1.1 based on at least four chains to state that the chains were sufficiently close to convergence for all parameters $\theta$. This corresponds to a situation where the variance of the parameters is lower between chains than within them indicating that all the chains have identified the same stationary distribution.

\subsection{Signal detection criteria}\label{sec:detection_criteria}

Throughout the analyses, we used the signal detection criteria of \citet{tuomi2012}. These criteria have recently been applied in e.g. \citet{anglada2013}, \citet{tuomi2013a}, \citet{tuomi2013b}, and \citet{tuomi2013c}, and they appear to be trustworthy in the sense that they are not particularly prone to false positives \citep{tuomi2012c}, and enable the detections of signals that cannot be found by using more traditional detection criteria based on false alarm probabilities (FAPs) in the power spectrum of the model residuals \citep{anglada2012b,anglada2013,tuomi2013b}. As an example, we refer to \citet{tuomi2013c}, who independently discovered the same three planet candidates around GJ 163 as \citet{bonfils2013b} with only $\sim$ 35\% of the data. This indicates that the Bayesian signal detection criteria indeed are very sensitive and robust ones in detecting weak signals of low-mass planets in radial velocity data.

The first criterion is that a model with $k+1$ signals, denoted as $\mathcal{M}_{k+1}$, has a posterior probability that is at least $s$ times greater than the corresponding probability of a model with only $k$ signals, i.e. $P(\mathcal{M}_{k+1} | m) \geq s P(\mathcal{M}_{k} | m)$. If this criterion is not satisfied, the $k+1$th signal has a considerable probability of being produced by noise instead of being a genuine periodicity in the data. When analysing the data with $k=0, 1, ...$ it might happen that $P(\mathcal{M}_{k} | m) < s P(\mathcal{M}_{k-1} | m)$ but that $P(\mathcal{M}_{k+1} | m) \geq s P(\mathcal{M}_{k} | m)$. This means that it cannot be said there are significantly $k$ signals, and if the model $\mathcal{M}_{k+1}$ was not tested, the conclusion would be that there are only $k-1$ signals in the data. Therefore, if we observe putative probability maxima in the parameter space for model $\mathcal{M}_{k}$ that do not satisfy the detection criterion, we typically analyse the data with models $\mathcal{M}_{k+1}$ and $\mathcal{M}_{k+2}$ as well to see if the superposition of more than $k$ signals makes the detection of only $k$ signals impossible.

We note that the probability threshold $s$ has to be chosen subjectively. A typical choice would be $s = 150$ because it has been interpreted as corresponding to \emph{strong evidence} and recommended by e.g. \citet{kass1995} based on the arguments of \citet{jeffreys1961}, although it has been argued that a more conservative threshold of 1000 should be chosen \citep{evett1991}. This threshold of $s = 150$ has been applied in analyses of radial velocity data succesfully \citep[e.g.][]{feroz2011,gregory2011,tuomi2012,tuomi2013a,tuomi2013b}. However, this choice might result in overinterpretation of the significance when the statistical model used to describe the data does not represent the data well. For the purpose of population studies such as the current work, one should be more cautious than usual in accepting new planet candidates to avoid the contamination of occurrence rate estimates by false positives. Therefore, we remain cautious about the interpretation of signals that are detected with low thresholds, and use a more conservative threshold of $s = 10^{4}$ as a requirement for a planet candidate.

The second criterion is related to the first one in the sense that it states that the radial velocity amplitude has to be statistically significantly (with e.g. 99\% level) greater than zero. If this was not the case, there would be a considerable probability that the amplitude of the signal was actually negligible and that the signal did not thus exist in reality. The third criterion states that the period of a signal has to be well-constrained from below and above to consider it a genuine periodicity. Again, the justification of this criterion is simple because if the period parameter was not well constrained, it could not be claimed that the corresponding variations in the data were indeed of periodic nature. We note that in practice, based on the analyses in the present work, a signal whose significance exceeds $s = 10^{4}$ is always well constrained in the parameter space. Conversely, if a given signal is constrained in the period and amplitude space, its significance always exceeds some $s > 1$, though not necessarily $s = 150$ or $s = 10^{4}$. This means that the detection criteria are complementary and robust in detecting low-amplitude signals.

To assess whether any given signal observed in the combined data set is supported by both data sets, we examine the residuals of each model using common weighted root-mean-square (RMS) statistics. If these statistics are decreased for both data sets when adding a signal to the model, we say that both data sets support the existence of the corresponding signal even when the signal cannot be detected in both data sets independently. This choice also serves as a test of whether a given weak signal could be a spurious one generated by noise and the correspondingly insufficiently accurate model. For a given signal to be a genuine one of stellar origin (as opposed to a spurious signal of instrumental origin) -- possibly caused by the Doppler variations induced by an orbiting planet -- it should result in a decrease in the RMS estimates of both data sets, although this is also subject to chance and does not necessarily hold when there are only few measurements.

Finally, to make the detections as robust as possible, we set the following criteria for signals to be able to call them planet candidates. First, we require that such signals are detected with a choice of $s = 10^{4}$ and can be concluded to be present in the corresponding combined data set(s) very significantly. If counterparts of these signals cannot be found in the activity indices calculated from the UVES and HARPS spectra, they are considered planet candidates and we refer to them according to the standard nomenclature by assigning letters b, c, ... to them. If, however, (1) the corresponding signals are detected \emph{only} with a significance of $s < 10^{4}$, (2) if there is data available from only one spectrograph such that the existence of a signal is not supported by an independent data set because such a set is not available, or (3) if a signal is present only in one of the data sets (in the sense that the RMS of the other does not decrease when adding the signal to the model) that dominates the posterior density given the combined data because of the low number or poor quality of the measurements in the other set, we do not call them candidate planets.

We note that if the posterior density consists of several other (local) maxima in the parameter space at several different periods with reasonably high probabilities (e.g. with $n$ local maxima at $\theta^{i}_{\rm L}, i = 1, ..., n$, that have $\pi(\theta^{i}_{\rm L} | m) \geq 0.01 \pi(\theta_{\rm MAP} | m)$ for all $i$, where $\theta_{\rm MAP}$ is the MAP estimate), it is possible that there are also several significant solutions in the period space that satisfy the detection criteria. Such a situation is hard to interpret reliably, but can arise for two main reasons. First, all the significant periodicities can correspond to genuine periodic signals in the velocity data -- several planetary signals in the data may have similar amplitudes and be detected roughly equally confidently \citep[e.g. GJ 667C whose Doppler data was found to contain six signals of roughly equal amplitude;][]{anglada2013}. Alternatively, such a situation might be representative of poor noise modelling where periodic/quasiperiodic features and/or correlations in the noise are falsely interpreted as genuine signals. In such cases, we analyse the data by increasing $k$ further but report the solution corresponding to the global maximum if there is no strong evidence in favour of additional signals in the sense of our detection criteria.

\subsection{Search for significant periodicities}\label{sec:search_techniques}

We performed the searches for significant periodicities in the data sets in several stages. First, we obtained a sample from the posterior density of the model without any Keplerian signals. In the second step, we sampled the posterior density of the model with $k=1$ by using tempered chains such that the posterior density was raised to a power of $\beta \in (0,1)$. In particular, we used $\beta =$ 0.3 - 0.8 to find the most promising areas in the period space because such a choice of $\beta$ enabled a rapid search of the whole period space $[T_{0}, T_{\rm obs}$], where $T_{\rm obs}$ is the data baseline and we selected $T_{0} =$ 1 day. However, if there were clear indications of probability maxima in excess of a period equal to $T_{\rm obs}$, we increased the upper limit to $2T_{\rm obs}$. We did not use the adaptive Metropolis algorithm \citep{haario2001} but the standard Metropolis-Hastings version \citep{metropolis1953,hastings1970} in these periodicity searches to prevent the proposal density from adapting to the possibly very narrow global maximum in the posterior density because the purpose of these samplings was to identify the positions of the most significant probability maxima in the parameter (period) space and not to enable the chain to adapt to one of them. 

For most data sets, we started by a test sampling with a ``cold'' chain, i.e. a chain with $\beta = 1$. When the number of measurements was lower than $\sim$ 15, these cold samplings were sufficient in exploring the whole period-space because typically there were no significant maxima in the parameter space that would have prevented the chain from exploring the whole period space rapidly. When such cold samplings were not possible due to an abundance of reasonably high maxima in the parameter space and thus poor mixing of the Markov chains in the period space, we used tempered chains by setting $\beta = 0.3$ to 0.5, depending on the number of measurements. The lowest values of 0.3 were used for the largest data sets of GJ 699 ($N_{\rm UVES} = 226$), GJ 551 ($N_{\rm UVES} = 229$), and GJ 433 ($N_{\rm UVES} = 166$). These samplings resulted in estimates for the positions of the most significant maxima in the period space. However, it must be noted that using tempered samplings corresponds actually to using different prior and likelihood models in the analysis such that $\pi^{\beta}(\theta)$ and $l^{\beta}(m | \theta)$ are used as a prior density and likelihood function instead of $\pi(\theta)$ and $l(m | \theta)$, respectively. Therefore, the shape of the posterior density, as estimated based on tempered samplings, does not necessarily correspond to the actual posterior density. However, the positions of the maxima are unchanged.

Given rough estimates for the positions of the probability maxima in the period space based on tempered samplings, we started several cold samplings with initial values in the vicinity of the observed maxima to enable fast convergence. If one (or some) of these maxima corresponded to a significant periodic signal(s) in the sense of the detection criteria we described above (Section \ref{sec:detection_criteria}), we continued by increasing $k$ and by performing tempered samplings of the parameter space of a model with one more Keplerian signal. These samplings were performed by using cooler chains because after the strongest signal was accounted for by the model, samplings with $\beta =$ 0.6 - 0.8 were typically found to visit all areas of the period space of the additional signal with $\sim$ few 10$^{6}$ chain members.

Finally, we obtained samples from the posterior density of a model with $k+1$ Keplerian signals when the corresponding data was found to contain significant evidence in favour of only $k$ signals, i.e. that only $k$ signals satisfied the detection criteria. These samplings were performed to estimate which additional signals could be allowed by the data and which ones could thus be ruled out \citep[e.g.][]{tuomi2012}.

We note that some data sets have only two UVES measurements and/or less than two HARPS measurements. Therefore, because we used a linear trend in the model, we did not analyse UVES data sets with less than three mesurements. Furthermore, we did not analyse the combined set when the number of measurements in either UVES or HARPS set is less than two. This means that we might analyse an individual data set with three measurements and a combined one with four (two UVES and two HARPS) measurements. Even in such extreme cases, when the number of free parameters in the statistical model exceeds the number of measurements, we expect to be able to obtain meaningful results because we use informative prior densities and because the Bayesian statistical techniques we use take the principle of parsimony, and thus the possible effects of such overparameterisation, into account and do not rely on assumptions regarding the number of parameters or measurements.

\subsection{Example: GJ 229}

To demonstrate the practicality of our periodicity search technique, we show the analysis results in detail for the UVES velocities of GJ 229 because it is a typical target in the sample with a reasonably large number of measurements ($N_{\rm UVES} = 73$) that show evidence in favour of periodic variations. We started by analysing the data with the benchmark model with $k=0$.

\begin{figure}
\center
\includegraphics[angle=270, width=0.40\textwidth]{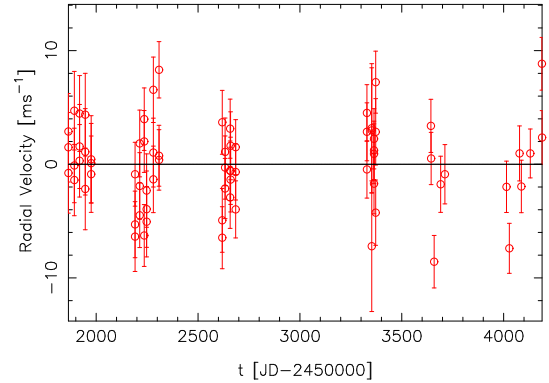}
\caption{Residuals of the benchmark model without Keplerian signals for the UVES data of GJ 229.}\label{fig:gj229_0p_resid}
\end{figure}

We found that the UVES data of GJ 229 contained a significant amount of correlation with $\phi_{\rm UVES} =$ 0.85 [0.45, 1.00]; evicence for a linear trend with $\dot{\gamma} =$ 1.26 [0.39, 2.17] ms$^{-1}$year$^{-1}$, small part of which is caused by secular acceleration\footnote{The secular acceleration was not subtracted from the UVES data of \citet{zechmeister2009} at this stage. However, we subtracted it when analysing the UVES data in combination with the HARPS velocities from which the secular acceleration was subtracted by the TERRA processing.} of 0.070 ms$^{-1}$year$^{-1}$ \citep{zechmeister2009}; and excess Gaussian white noise with $\sigma_{\rm UVES} =$ 2.51 [1.29, 3.74] ms$^{-1}$, which is rather low and implies that the star expresses very low levels of velocity variability and/or the UVES instrument uncertainties in the \citet{zechmeister2009} velocities might be overestimated. Removing the MAP trend and correlations, we plotted the resulting residuals in Fig. \ref{fig:gj229_0p_resid}. These residuals can indeed be described as ``flat'', which indicates that there cannot be periodic variations in these velocities with amplitudes greater than roughly 10 ms$^{-1}$. The RMS of the residuals is 3.61 ms$^{-1}$, which is considerably lower than that of the original velocities of 5.26 ms$^{-1}$.

\begin{figure}
\center
\includegraphics[angle=270, width=0.40\textwidth]{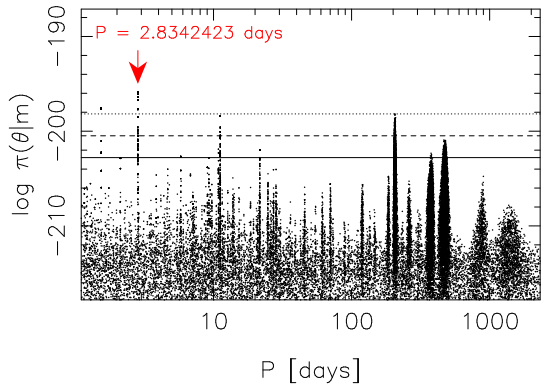}
\caption{Log-posterior density as a function of the log-period parameter of a Keplerian signal from a tempered sampling ($\beta = 0.3$) of the GJ 229 UVES data. The horizontal lines indicate the 10\% (dotted), 1\% (dashed), and 0.1\% (solid) probability thresholds with respect to the MAP value that is denoted by using the red arrow.}\label{fig:gj229_tempered}
\end{figure}

We performed a tempered search for periodicities, and obtained a sample from the posterior density corresponding to a choice of $\beta = 0.3$. We plotted an example of a corresponding sampling in Fig. \ref{fig:gj229_tempered}. As can be seen, the chain identifies a global maximum in the period space of the one-Keplerian model at a period of 2.8 days (red arrow). The chain also visits the whole period space between 1 day and $T_{\rm obs} = 2325$ days for GJ 229. The chain in Fig. \ref{fig:gj229_tempered} also shows that there are additional local maxima in the period space exceeding the 10\%, 1\% and/or 0.1\% probability levels (dotted, dashed, and solid horisontal lines, respectively) with respect to the global maximum at periods of roughly 1.5, 10, 200, and 450 days. The existence of such multiple maxima in the scaled posterior suggests that none of them can be confidently considered a solution and thus a periodic signal that is reliably detected in the data. Our samplings also suggest that there are likely no other considerable probability maxima in the period space, or that if they exist, they are so narrow that the chains are unlikely to visit them. We performed such samplings several times and obtained consistent results with no indications of additional maxima with similar posterior values.

Because of the candidate periodicities, we started cold chains with initial periods in the vicinity of the highest maxima (Fig. \ref{fig:gj229_tempered}) and were able to identify two periodicities that satisfied the detection criteria discussed above. These periodicities were found at periods of 1.5 and 206 days with amplitudes of 5.3 and 4.3 ms$^{-1}$ days, respectively, but it cannot be concluded that we have identified such signals confidently in the data because the posterior density has multiple maxima comparable to the posterior at the MAP estimate. Therefore, we can only conclude that if there are periodic signals in the UVES data, they are likely present at the periods corresponding to the maxima in the posterior density (Fig. \ref{fig:gj229_tempered}).

We performed samplings of the UVES data with a model containing two Keplerian signals, but could not find significant two-Keplerian solutions to the data.

It should be noted that the samples from the (scaled) posterior densities in Fig. \ref{fig:gj229_tempered} can be roughly interpreted in a similar manner as e.g. common Lomb-Scargle \citep{lomb1976,scargle1982} periodograms. This is because the plotted density represents the relative significances of the different periods and can thus be broadly interpreted according to the probabilities presented in the vertical axis showing the log-posterior values. However, the relative probabilities of the corresponding periodicities, and indeed whether they are significant enough to satisfy the detection criteria, can only be assessed by additional samplings and by seeing whether the periods can be constrained from above and below.

\begin{figure}
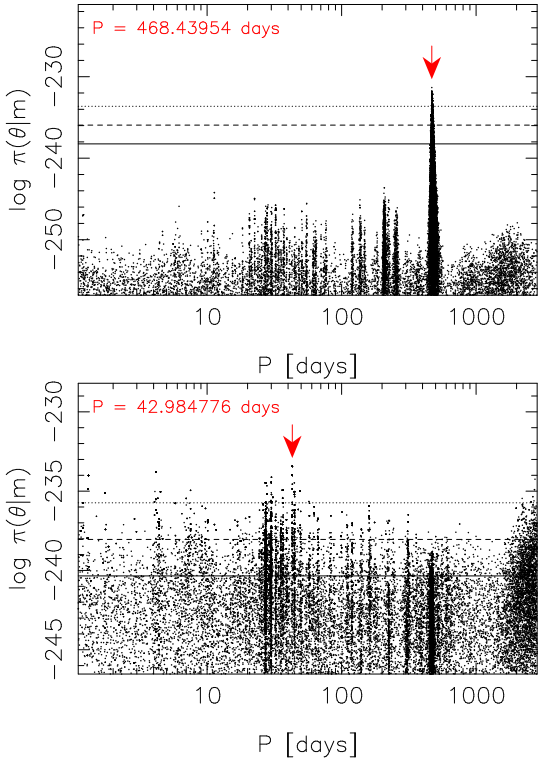

\center
\includegraphics[angle=270, width=0.40\textwidth]{rvdist01_GJ229b_psearch_b.ps}

\includegraphics[angle=270, width=0.40\textwidth]{rvdist02_GJ229b_psearch_c.ps}
\caption{As in Fig. \ref{fig:gj229_tempered} for the first signal in the combined HARPS and UVES data of GJ 229 (top panel), and for a cold sampling plotted as a function of the second signal of a two-Keplerian model indicating that there are no additional periodic signals (bottom panel).}\label{fig:GJ229_tempered_combined}
\end{figure}

\begin{figure}
\center
\includegraphics[angle=270, width=0.40\textwidth]{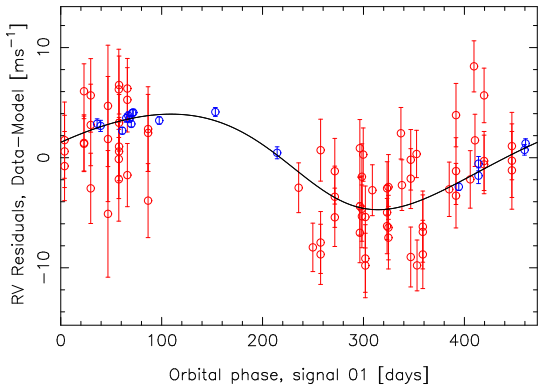}
\caption{Phase folded signal in the combined HARPS (blue) and UVES (red) data of GJ 229.}\label{fig:GJ229_curve}
\end{figure}

One last test for a significance of a signal is whether it is present in two independent data sets or not. According to our results, the putative signals in the UVES data at periods of 1.5 and 206 days fail this test as the HARPS data effectively rules them out. This can be seen by looking at the corresponding log-posterior density as a function of the period of a signal in the combined data (Fig. \ref{fig:GJ229_tempered_combined}, top panel). Accordingly, there is a strong and isolated global maximum at a period of roughly 470 days that satisfies all the detection criteria discussed above. Because this signal is clearly present as a local maximum in the UVES data alone, and because there are no comparable local maxima in the period space given the combined data, we conclude that there is a significant periodic signal in the combined HARPS and UVES velocities of GJ 229 at a period of 471 [459, 493] days with an amplitude of 3.83 [2.15, 5.57] ms$^{-1}$ that corresponds to a new and previously unknown planet candidate orbiting the star with a minimum mass of 32 [16, 49] M$_{\oplus}$. This signal can be called a candidate planet because its existence is supported by both data sets according to our requirements and because it is detected very confidently in the combined data. Furthermore, as we discuss in Section \ref{sec:activity}, the signal does not correspond to variations in the activity data of GJ 229. To demonstrate its significance, we have plotted the phase-folded signal in Fig. \ref{fig:GJ229_curve}. We note that there is no evidence in favour of a second signal in the combined data, as can also be seen in the bottom panel of Fig. \ref{fig:GJ229_tempered_combined} that does not have strong and isolated maxima in the period space.

\subsection{Detection probability and planet occurrence}\label{sec:occurrence_computations}

To obtain estimates of the underlying occurrence rate of planets in our sample, it is necessary to estimate the detection bias of the current data caused by the fact that radial velocity data sets are more sensitive to greater planetary masses and shorter orbital periods. It is also necessary to account for the quantity and quality of each data set, as well as their respective baselines. For this purpose, we calculated detection probabilities for the combined UVES and HARPS data sets of the sample by using samplings of the parameter space as shown in e.g. the bottom panel of Fig. \ref{fig:GJ229_tempered_combined}.

First, we estimate that in any given data set (the $i$th data set), the observed number of signals in a given period interval $\Delta P$ and minimum mass interval $\Delta M$, that we denote as $\Delta_{\rm P,M}$ for short, can be written as $f_{\rm obs,i}(\Delta_{\rm P,M}) = f_{\rm occ,i} (\Delta_{\rm P,M}) p_{i}(\Delta_{\rm P,M})$, where $f_{\rm occ,i}$ is the number of planets orbiting the $i$th star in the sample and $p_{i}$ is the detectability function that indicates whether a planet with parameters in the interval $\Delta_{\rm P,M}$ can be detected in the data set $m_{i}$. However, while $f_{\rm obs,i}$ is easy to obtain as it is directly the number of planets we detect in the respective mass-period interval in data set $m_{i}$, $p_{i}$ is more difficult to calculate.

To estimate $p_{i}$, we use the posterior samplings of a model with $k+1$ Keplerian signals when there are only $k$ signals in the data set. The sample drawn by using the posterior sampling enables us to reconstruct the areas of the parameter space where the parameters describing the hypothetical $k+1$th signal visited. Because the signal cannot be detected in the sense that its amplitude and period could be constrained, the chain visits all areas in the period space allowed by the chosen range of the period (from one day to data baseline). Moreover, the chain visits, given sufficiently good mixing properties such that the chain visits all relevant areas in the parameter space frequently, all amplitudes at all periods that are \emph{allowed} by the data, i.e. amplitudes that are so low that they cannot be ruled out by the data because the likelihood function has reasonably high values. The chain does not visit amplitudes in excess of some limiting amplitude at each period because they would correspond to such low likelihoods that planetary signals with such amplitudes are effectively ruled out. Therefore, the sampling yields the areas of parameter space where there \emph{could} be signals. The rest of the parameter space is thus the area where there are (very likely) no additional signals based on the data.

\begin{figure}
\center
\includegraphics[angle=270, width=0.45\textwidth]{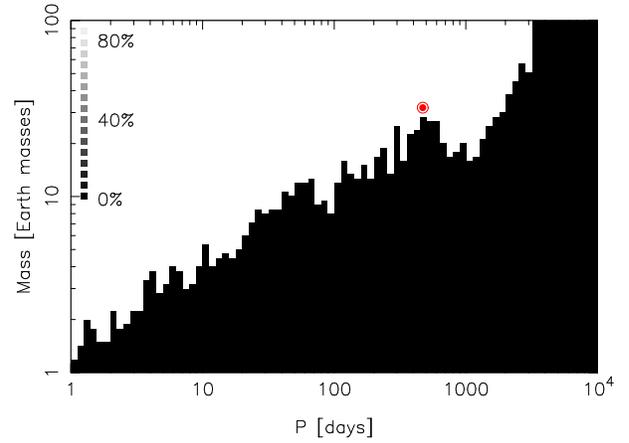}
\caption{Detection threshold of the combined data of GJ 229. The white area corresponds to parameter values in the mass-period space where signals could be detected in the data (detection probability of unity) whereas the black area shows a corresponding mass-period space where signals cannot be detected (negligible detection probability). The candidate GJ 229 b is shown as a red circled dot that is in the former area because it was detected.}\label{fig:GJ229_threshold}
\end{figure}

We demonstrate this by using GJ 229 as an example. It has one candidate and thus we use a model with two of them to estimate the detectability function. In Fig. \ref{fig:GJ229_threshold}, the black area shows the subset of the mass-period space where the mass and period parameters of the second signal visited during the Markov chain samplings. This area thus corresponds to signals that could exist in the data but that cannot be detected because we did not find a second signal in the GJ 229 data. The white area corresponds to mass-period space where the chains did not visit and thus rule out planetary signals because the likelihood function is so low in these areas that the chances of a planet existing in the white area can be approximated to be negligible. The planet candidate GJ 229 b is clearly above the threshold as it should because it was detected. We note that while this planet candidate has parameters close to the threshold, it is still several M$_{\oplus}$ heavier than the minimum-mass planet that could be detected at that period. In fact, if it was much more above the threshold, it is likely that its existence would already have been reported based on UVES or HARPS data sets alone.

We thus estimate that planet candidates could have been detected in the areas where the Markov chains did not visit (white areas in Fig. \ref{fig:GJ229_threshold}). This is the case because they correspond to the complement of the area where planets could not be detected. In reality, however, the threshold is not so strict because the detection probability is a (continuous) function of the parameters. We do not attempt to estimate this function more accurately in this work and use the ''step-function`` for each data set as shown in Fig. \ref{fig:GJ229_threshold} for GJ 229 as a first-order approximation.

Assuming that there are some areas in the parameter space where the chains did not visit due to e.g. poor sampling even though they should have because these areas have sufficiently high likelihoods, we would effectively overestimate the detection probabilities and thus underestimate the occurrence rates. Thich means that our occurrence rate estimates are lower limits, although unlikely to be considerably lower to bias the results significantly given the rather large uncertainties due to the low number of planets in the sample. We approximate the detectability by setting the function $\hat{p}_{i} = 1- p_{i}$ equal to unity in the areas where the Markov chains did visit, and equal to zero where they did not.

We express the detected frequency of planets for the whole sample of stars, with data sets $m_{1}, ..., m_{N}$, by summing the number of observed signals $f_{\rm obs,i}$ of all the data sets and by assuming that the occurrence rate is common for all stars in the sample such that $f_{\rm occ} = f_{\rm occ,i}$ for all $i$ (in units of planets per star). Thus we obtain
\begin{eqnarray}\label{eq:detection}
  && f_{\rm obs}(\Delta_{\rm P,M}) = \sum_{i=1}^{N} f_{\rm obs,i}(\Delta_{\rm P,M}) \nonumber\\
  && = f_{\rm occ} (\Delta_{\rm P,M}) \Bigg[ N - \sum_{i=1}^{N} \hat{p}_{i}(\Delta_{\rm P,M}) \Bigg] ,
\end{eqnarray}
which implies a simple way of calculating the occurrence rate $f_{\rm occ}$ for the whole sample. In this equation, the term in square brackets on the right hand side is, when divided by $N$, the detection probability function of our sample that approximates the probability of being able to detect planets in the interval $\Delta_{\rm P,M}$. Thanks to our samplings, we could estimate this function rather accurately - typically by using a 100$\times$100 grid in the log-parameter space ranging from a minimum period of 1 day to a maximum of 10$^{4}$ days and from a minimum mass of 1 M$_{\oplus}$ to a maximum of 100 M$_{\oplus}$. Such a fine grid was not a practical choice for estimating the occurrence rates, because to obtain any meaningful estimates at all, it would be desirable to have at least one planetary signal in most grid points. For this reason, when calculating occurrence rates, we divided the interval into a 4$\times$4 grid.

When estimating the uncertainties for the occurrence rates in a given interval $\Delta_{\rm P,M}$, we use rather conservatively the lowest and highest detection probabilities (obtained by using the finer grid) in this interval to calculate lower and upper limits, respectively.

\section{The sample and Keplerian signals}\label{sec:UVES}

The sample of M dwarfs for which the UVES velocities were published in \citet{zechmeister2009} contains a collection of nearby stars including the two nearest M dwarfs, namely GJ 551 and GJ 699 that are the nearest and fourth nearest stars to the Sun, respectively. We have listed the stars in this sample in Table \ref{tab:target_properties} together with their estimated physical properties. While the parallaxes are obtained from Hipparcos \citep{vanleeuwen2007} and the mass estimates from \citet{zechmeister2009}, we estimated the effective temperatures and luminosities by using the empirical relations of \citet{casagrande2008} and \citet{boyajian2012}. Based on the estimated luminosities and effective temperatures, we have also calculated the approximate inner and outer edges of the stellar habitable zones according to the equations of \citet{kopparapu2013} and listed them in Table \ref{tab:target_properties}. We note that the stars in the sample have also been extensively searched for brown dwarf companions. For instance, \citet{dieterich2012} reported that $0.0^{+3.5}_{-0.0}$ \% occurrence rate of L  and T companions in separations between 10--70 AU. Furthermore, the sample stars show no evidence for warm \citep[e.g.][]{avenhaus2012} or cold \citep[e.g.][]{lestrade2009} circumstellar material.

\begin{table*}
\caption{Target stars, their Hipparcos parallaxes, and physical properties together with the estimated inner and outer edges of the stellar habitable zones based on \citet{kopparapu2013}.}\label{tab:target_properties}
\begin{minipage}{\textwidth}
\begin{center}
\begin{tabular}{llcccccccc}
\hline \hline
Star & SC\footnote{Spectral types as reported in \citet{koen2010} unless mentioned otherwise: (1) \citet{endl2006}; (2) \citet{torres2006}; (3) \citet{jenkins2009}; (4) \citet{reid2004}; (5) \citet{gray2006}; (6) \citet{montes2001}.} & $\pi$\footnote{Hipparcos parallaxes \citep{vanleeuwen2007}.} & $M_{\star}$\footnote{Estimates from \citet{zechmeister2009}.} & $L_{\star}$ & $T_{\rm eff}$ & HZ (inner) & HZ (outer) & Notes\footnote{(MC) evidence for a massive substellar companion whose orbit cannot be constrained; (PC) planetary companions \citep{delfosse2012}; (DI) directly imaged brown dwarf companion \citep{nakajima1995}; (BD) brown dwarf companion \citep{kurster2008}.}. \\
& & [mas] & [M$_{\odot}$] & [L$_{\odot}$] & [K] & [AU] & [AU] \\
\hline
GJ 1 & M1.5 V$^{\rm }$ & 230.42 $\pm$0.90 & 0.45 & 0.025 & 3527 & 0.17 & 0.32 \\
GJ 27.1 & M0.5 V$^{\rm }$ & 41.69 $\pm$2.80 & 0.53 & 0.026 & 3542 & 0.17 & 0.33 \\
GJ 118 & M2.5 V$^{\rm }$ & 85.87 $\pm$1.99 & 0.36 & 0.0093 & 3292 & 0.10 & 0.20 \\
GJ 160.2 & M0 V\footnote{\citet{zechmeister2009} adopt a spectral type of M0 but more recently \citet{koen2010} have reported a spectral type of K7.} & 43.25 $\pm$1.61 & 0.69 & 0.16 & 4347 & 0.41 & 0.76 \\
GJ 173 & M1.5 V$^{\rm 4}$ & 90.10 $\pm$1.74 & 0.48 & 0.018 & 3438 & 0.14 & 0.27 \\
GJ 180 & M2 V$^{\rm }$ & 85.52 $\pm$2.40 & 0.43 & 0.013 & 3371 & 0.12 & 0.24 \\
GJ 190 & M3.5 V$^{\rm }$ & 107.85 $\pm$2.10 & 0.44 & 0.0042 & 3143 & 0.07 & 0.14 \\
GJ 218 & M1.5 V$^{\rm }$ & 66.54 $\pm$1.43 & 0.50 & 0.019 & 3449 & 0.15 & 0.28 \\
GJ 229 & M1/M2 V$^{\rm 3}$ & 173.81 $\pm$0.99 & 0.58 & 0.028 & 3564 & 0.18 & 0.34 & DI \\
GJ 263 & M3.5 V$^{\rm 4}$ & 62.41 $\pm$3.16 & 0.55 & 0.0054 & 3187 & 0.08 & 0.15 \\
GJ 357 & M2.5 V$^{\rm 5}$ & 110.82 $\pm$1.92 & 0.37 & 0.011 & 3328 & 0.11 & 0.22 \\
GJ 377 & M3 V$^{\rm }$ & 61.39 $\pm$2.55 & 0.52 & 0.0081 & 3264 & 0.10 & 0.19 \\
GJ 422 & M3.5 V$^{\rm }$ & 78.91 $\pm$2.60 & 0.35 & 0.011 & 3323 & 0.11 & 0.21 \\
GJ 433 & M1.5 V$^{\rm }$ & 112.58 $\pm$1.44 & 0.48 & 0.020 & 3472 & 0.15 & 0.29 & PC \\
GJ 477 & M1 V$^{\rm }$ & 52.67 $\pm$3.05 & 0.54 & 0.018 & 3441 & 0.14 & 0.27 & MC \\
GJ 510 & M1 V$^{\rm }$ & 59.72 $\pm$2.43 & 0.49 & 0.019 & 3460 & 0.15 & 0.28 \\
GJ 551 & M6 V$^{\rm 2}$ & 771.64 $\pm$2.60 & 0.12 & 0.0022 & 3042 & 0.05 & 0.10 \\
GJ 620 & M0 V$^{\rm 5}$ & 60.83 $\pm$2.06 & 0.61 & 0.038 & 3661 & 0.21 & 0.40 \\
GJ 637 & M$^{\rm }$ & 62.97 $\pm$1.99 & 0.41 & 0.022 & 3492 & 0.16 & 0.30 \\
GJ 682 & M3.5 V$^{\rm }$ & 196.90 $\pm$2.15 & 0.27 & 0.0020 & 3028 & 0.05 & 0.09 \\
GJ 699 & M4 V$^{\rm 3}$ & 548.31 $\pm$1.51 & 0.16 & 0.0021 & 3034 & 0.05 & 0.10 \\
GJ 739 & M$^{\rm }$ & 70.95 $\pm$2.56 & 0.45 & 0.012 & 3352 & 0.12 & 0.23 \\
GJ 817 & M1 V$^{\rm }$ & 52.16 $\pm$2.92 & 0.43 & 0.028 & 3561 & 0.18 & 0.34 \\
GJ 821 & M1 V$^{\rm }$ & 82.18 $\pm$2.17 & 0.44 & 0.023 & 3512 & 0.16 & 0.31 \\
GJ 842 & M0.5 V$^{\rm }$ & 83.43 $\pm$1.77 & 0.58 & 0.031 & 3590 & 0.19 & 0.36 \\
GJ 855 & M1 V$^{\rm }$ & 52.22 $\pm$2.17 & 0.60 & 0.028 & 3567 & 0.18 & 0.34 \\
GJ 891 & M2V$^{\rm 5}$ & 62.17 $\pm$3.27 & 0.35 & 0.015 & 3404 & 0.13 & 0.26 \\
GJ 911 & M0 V$^{\rm 1}$ & 41.22 $\pm$2.64 & 0.63 & 0.047 & 3731 & 0.23 & 0.44 \\
GJ 1009 & M1.5 V$^{\rm }$ & 55.62 $\pm$2.32 & 0.56 & 0.015 & 3396 & 0.13 & 0.25 \\
GJ 1046 & M2.5 V$^{\rm 5}$ & 71.06 $\pm$3.23 & 0.40 & 0.0089 & 3283 & 0.10 & 0.20 & BD \\
GJ 1100 & M0 V$^{\rm 4}$ & 34.57 $\pm$2.79 & 0.57 & 0.045 & 3717 & 0.23 & 0.43 \\
GJ 3020 & M2.5 V$^{\rm }$ & 43.89 $\pm$4.39 & 0.62 & 0.012 & 3347 & 0.12 & 0.23 & MC \\
GJ 3082 & M0 V$^{\rm }$ & 60.38 $\pm$1.81 & 0.47 & 0.022 & 3497 & 0.16 & 0.30 \\
GJ 3098 & M1.5 V$^{\rm 5}$ & 55.98 $\pm$1.91 & 0.50 & 0.024 & 3518 & 0.17 & 0.32 \\
GJ 3671 & M0 V$^{\rm }$ & 56.38 $\pm$2.04 & 0.50 & 0.030 & 3584 & 0.19 & 0.35 \\
GJ 3759 & M1 V$^{\rm 5}$ & 58.94 $\pm$2.40 & 0.49 & 0.026 & 3545 & 0.17 & 0.33 \\
GJ 3916 & M2.5 V$^{\rm 5}$ & 66.21 $\pm$3.18 & 0.49 & 0.0091 & 3288 & 0.10 & 0.20 & MC \\
GJ 3973 & M1.5 V$^{\rm 5}$ & 54.86 $\pm$2.18 & 0.54 & 0.024 & 3524 & 0.17 & 0.32 \\
GJ 4106\footnote{The luminosity and $T_{\rm eff}$ appear to be inconsistent with the spectral class of M2.} & M2 V$^{\rm 6}$ & 9.05 $\pm$3.70 & 0.55 & 0.20 & 4552 & 0.47 & 0.85 \\
GJ 4293 & M$^{\rm }$ & 39.90 $\pm$3.04 & 0.57 & 0.0023 & 3044 & 0.05 & 0.10 \\
HG 7-15 & M1 V$^{\rm 1}$ & 26.80 $\pm$2.05 & 0.78 & 0.011 & 3323 & 0.11 & 0.21 \\
\hline \hline
\end{tabular}
\end{center}
\end{minipage}
\end{table*}

We obtained the HARPS-TERRA (\emph{Template Enhanced Radial velocity Re-analysis Application}) velocities from the publicly available spectra\footnote{We also obtained three additional HARPS spectra of GJ 699 to increase the baseline of the HARPS data.} by using the data processing algorithms of \citet{anglada2012c}. We chose to use the TERRA velocities instead of the commonly used HARPS-CCF (cross-correlation function) velocities because of their lower scatter and therefore better precision for M dwarfs \citep{anglada2012c,anglada2012b,anglada2013,tuomi2013c}. While an abundance of such velocities could not be obtained for every star because a large fraction of the stars in our sample have not been primary targets of the HARPS-GTO survey \citep{bonfils2013}, there were still several stars for which the available HARPS data could be readily expected to provide better constraints for the possible planet candidates orbiting them due to its high precision, or help disputing the existence of putative signals in the UVES data as false positives. The numbers of HARPS measurements and the corresponding data baselines are listed in Table \ref{tab:UVES_signals} and we have tabulated the corresponding HARPS-TERRA velocities in the Appendix \ref{sec:velocities}.

\begin{table*}
\caption{Properties of the UVES and HARPS data sets in terms of numbers of measurements ($N_{\rm UVES}$, $N_{\rm HARPS}$) and data baselines ($\Delta_{\rm UVES}$, $\Delta_{\rm HARPS}$). Log-Bayesian evidence ratios, or Bayes factors ($\ln B_{i+1,i} = \ln P(m | \mathcal{M}_{i+1}) - \ln P(m | \mathcal{M}_{i})$), are shown when there is evidence in favour of at least one Keplerian signal according to our detection criteria. GJ 190 and GJ 263 are not in the table because they only had two velocity measurements. The four stars with evidence of massive companions in the radial velocity data are also not shown.}\label{tab:UVES_signals}
\begin{minipage}{\textwidth}
\begin{center}
\begin{tabular}{lccrrcccccccccl}
\hline \hline
Star & $N_{\rm UVES}$ & $N_{\rm HARPS}$ & $\Delta_{\rm UVES}$ & $\Delta_{\rm HARPS}$ & $\ln B_{1,0}$ & $\ln B_{2,1}$ & $k$ & Notes\footnote{(T) significant linear trend in the combined data; (D) solution consistent with the signals reported by \citet{delfosse2012}.}.\\
& & & [days] & [days] \\
\hline
GJ 1 & 37 & 44 & 2151.0 & 1843.0 & & & 0 &  \\
GJ 27.1 & 62 & 50 & 2152.1 & 437.0 & 14.3 & & 1 &  \\
GJ 118 & 56 & 14 & 2265.8 & 8.1 & & & 0 &  \\
GJ 160.2 & 100 & 7 & 2324.7 & 1894.9 & 12.0 & & 1 &  \\
GJ 173 & 12 & 5 & 896.7 & 3.9 & & & 0 &  \\
GJ 180 & 56 & 31 & 2324.7 & 351.1 & 16.4 & 10.2 & 2 &  \\
GJ 218 & 9 & 9 & 895.7 & 9.0 & & & 0 &  \\
GJ 229 & 73 & 17 & 2324.7 & 1724.2 & 18.8 & & 1 & T \\
GJ 357 & 70 & 5 & 2320.8 & 828.9 & & & 0 &  \\
GJ 377 & 14 & 16 & 1089.0 & 210.8 & & & 0 &  \\
GJ 422 & 24 & 25 & 1111.8 & 2281.9 & 10.2 & & 1 & T \\
GJ 433 & 166 & 62 & 2553.9 & 2244.9 & 26.9 & 12.2 & 2 & D \\
GJ 510 & 38 & 9 & 1114.8 & 94.7 & & & 0 & T \\
GJ 551 & 229 & 27 & 2555.0 & 1734.2 & & & 0 & T \\
GJ 620 & 5 & 11 & 421.7 & 115.6 & & & 0 &  \\
GJ 637 & 39 & 8 & 1098.9 & 99.8 & & & 0 &  \\
GJ 682 & 49 & 12 & 1134.0 & 1508.0 & 9.1 & 16.7 & 2 & T \\
GJ 699 & 226 & 25 & 2357.8 & 1869.9 & & & 0 & \\
GJ 739 & 49 & 2 & 1062.0 & 2.1 & & & 0 &  \\
GJ 817 & 49 & -- & 1550.8 & -- & & & 0 &  \\
GJ 821 & 106 & 5 & 1515.8 & 22.9 & & & 0 &  \\
GJ 842 & 17 & 7 & 925.7 & 19.0 & & & 0 &  \\
GJ 855 & 40 & 10 & 1560.8 & 1117.0 & & & 0 &  \\
GJ 891 & 46 & -- & 2178.0 & -- & & & 0 &  \\
GJ 911 & 27 & 3 & 2136.2 & 1456 & & & 0 &  \\
GJ 1009 & 34 & 12 & 2177.0 & 8.2 & & & 0 &  \\
GJ 1100 & 12 & 10 & 896.8 & 9.0 & & & 0 &  \\
GJ 3082 & 10 & -- & 760.8 & -- & & & 0 &  \\
GJ 3098 & 9 & -- & 732.8 & -- & & & 0 &  \\
GJ 3671 & 12 & 2 & 1090.0 & 17.8 & & & 0 &  \\
GJ 3759 & 11 & 11 & 1080 & 99.8 & & & 0 &  \\
GJ 3973 & 5 & 7 & 419.7 & 95.8 & & & 0 &  \\
GJ 4106 & 5 & -- & 396.0 & -- & & & 0 &  \\
GJ 4293 & 14 & -- & 874.6 & -- & & & 0 &  \\
HG 7-15 & 33 & -- & 416.8 & -- & & & 0 &  \\
\hline \hline
\end{tabular}
\end{center}
\end{minipage}
\end{table*}

We present the results of our comparisons of models containing $k=0, 1, 2$ Keplerian signals in Table \ref{tab:UVES_signals} by presenting the numbers of favoured signals in the combined data sets together with the numbers of measurements and the baselines of the UVES and HARPS data sets. According to these results, there are 10 significant signals in the combined data that satisfy our detection criteria and are supported by both data sets according to our requirements.

When publishing the analysis results of the same UVES datasets, \citet{zechmeister2009} did not report detections of planet candidates. According to our periodogram analyses with the standard Lomb-Scargle periodogram \citep{lomb1976,scargle1982} of the UVES data, this conclusion is indeed justified because, excluding the massive stellar or substellar companions\footnote{Apparent as variability of the order of few hundred ms$^{-1}$, although the corresponding orbits cannot be constrained.} around GJ 477, GJ 1046, GJ 3020, and GJ 3916; and excluding GJ 551 and GJ 699 for which \citet{zechmeister2009} reported signals caused by data sampling and activity, none of the UVES data sets had significant powers in their Lomb-Scargle periodograms corresponding to low-mass planetary companions. This inability to detect the signals of the planet candidates we report is not surprising because all these signals have very low amplitudes whose detection is difficult without a large number of measurements and high precision.

Interestingly, unlike \citet{zechmeister2009}, we could not find the 44-day signal they reported in the velocities of the GJ 699. We believe the reason is that this signal is likely caused by activity-related phenomena \citep{zechmeister2009} and is therefore quasiperiodic and/or time-dependent and explained rather well by correlations in the UVES measurements. Indeed, the estimate for the parameter $\phi_{\rm UVES}$ was found to be 0.81 [0.62, 0.99], which implies a considerable amount of correlation in the UVES data that could -- when coupled with data sampling -- give rise to significant but spurious powers in periodogram when not accounted for.

In addition to the signals we find, five targets in our sample show linear acceleration that is not consistent with zero suggesting the existence of yet unknown long-period substellar companions (Table \ref{tab:UVES_signals}).

\subsection{Analysis of activity indicators}\label{sec:activity}

To determine whether the signals we observe in the UVES radial velocities are caused by Doppler fingerprints of planetary and/or substellar companions or periodic or aperiodic and/or quasiperiodic phenomena related to the stellar activities, we analysed the line bisector (BIS) spans as obtained from the UVES spectra available in the ESO archive. BIS values indicate the line asymmetry and can be used as a signature of variations caused by stellar activity \citep[e.g.][]{queloz2001,boisse2011}. For most stars for which we observed radial velocity signals, the correlation coefficients between the velocities and BIS values were between -0.1 and 0.1, which implies no significant correlation. However, GJ 842 showed positive correlation coefficient of 0.43 between the radial velocities and BIS values that appears significant although the corresponding data set only contained 16 data points\footnote{We could only obtain 16 spectra for GJ 842 because the ESO archive did not contain calibration frames although there are 17 radial velocities in the \citet{zechmeister2009} data set (see also Table \ref{tab:UVES_signals}).}.

We also obtained the BIS values from the HARPS spectra and tested whether there were correlations between these values and the radial velocities. None of the data sets showed significant correlations between the BIS values and velocities, which indicates that the variations in the velocity data are unlikely to have been induced by activity-related phenomena of the stellar surface such as starspots and/or active regions. The typical correlation coeficients were found to be between -0.1 and 0.1, which is consistent with the interpretation that the signals we observe are genuine Doppler signatures of planetary nature.

\subsection{UVES false positives}

The UVES data contained some signals that were not supported by the HARPS velocities. This means that our benchmark model might not be optimal in the sense that features in the UVES measurement noise, or indeed biases, instrument-related variations, stability problems, etc. can produce variations that are interpreted as Keplerian signals. Furthermore, according to our analyses, even when we do detect a signal in the UVES data, the period space is typically littered by several other almost equally high local maxima that should be interpreted as alternative solutions before they can be ruled out. Such a multimodality occurs e.g. in the UVES data of GJ 160.2, GJ 229 (see Fig. \ref{fig:gj229_tempered}), GJ 377, and GJ 27.1.

We identified only two false positives in the UVES data that were essentially ruled out by the HARPS velocities. The first one of these is a 15-day periodicity in the 14 UVES velocity measurements of GJ 377. This signal increases the model probability with respect to the zero-Keplerian model by a factor of 700, which does not exceed our detection threshold of $10^4$. In this case, the HARPS data did not confirm the existence of the signal but ruled it out rather confidently. This weak evidence for a signal in the UVES data was also suspect because the period parameter of the signal had local probability maxima around 2, 5, 50-200, and 700 days that exceeded the 0.1\% or even 1.0\% probability thresholds of the global maximum. We interpret this result as hints of a signal in the UVES data whose exact period or amplitude cannot be determined due to the low number of measurements. Instead, in the combined dataset the global maximum of the period parameter was found at a period of roughly 80 days but it did not satisfy the detection criteria.

The other example is provided by the solution to the UVES data of GJ 357 that consists of two periodicities at 5 and 26 days, respectively. The significances of these signals decrease below the detection threshold with only five HARPS velocities. This means that -- even though there is not enough HARPS data to constrain any signals in the combined data set -- some of the local maxima in the UVES data can be ruled out\footnote{Here ''ruled out`` cannot be interpreted very strictly, as ruling out the existence of a signal is much harder in general than detecting one. }, but others might actually correspond to signals whose existence could be verified when future high-precision data becomes available.

We note that the iodine cell used in the UVES to obtain precise reference lines to the spectra can give rise to differences with respect to the ThAr method used in the HARPS. This means that some of the spurious signals and/or spurious local maxima that are unrelated to sampling in the UVES data can arise from small-scale instability in the UVES instrument. The same argument applies to the HARPS velocities as well, although HARPS has much greater long-term stability and precision. However, at the moment it is not possible to distinguish whether the signals with low significances that are interpreted as false positives in the UVES data because they do not correspond to global maxima in the combined data are caused by such instability, periodic or quasiperiodic features in the noise, periodicities related to the stellar activity and magnetic phenomena, and/or genuine Doppler signatures caused by low-mass planets. Ideally, signals arising from instrumentation can be ruled out by obtaining support for them from two data sets, and those caused by stellar surface phenomena can in principle be ruled out by showing that they or their primary harmonics do not have any couterparts in the activity-indices of the two sets. However, any remaining signals can still be caused by other unknown sources of periodic variation.

\section{Planets around M dwarfs}\label{sec:planet_statistics}

\begin{figure*}
\center
\includegraphics[angle=270, width=0.8\textwidth]{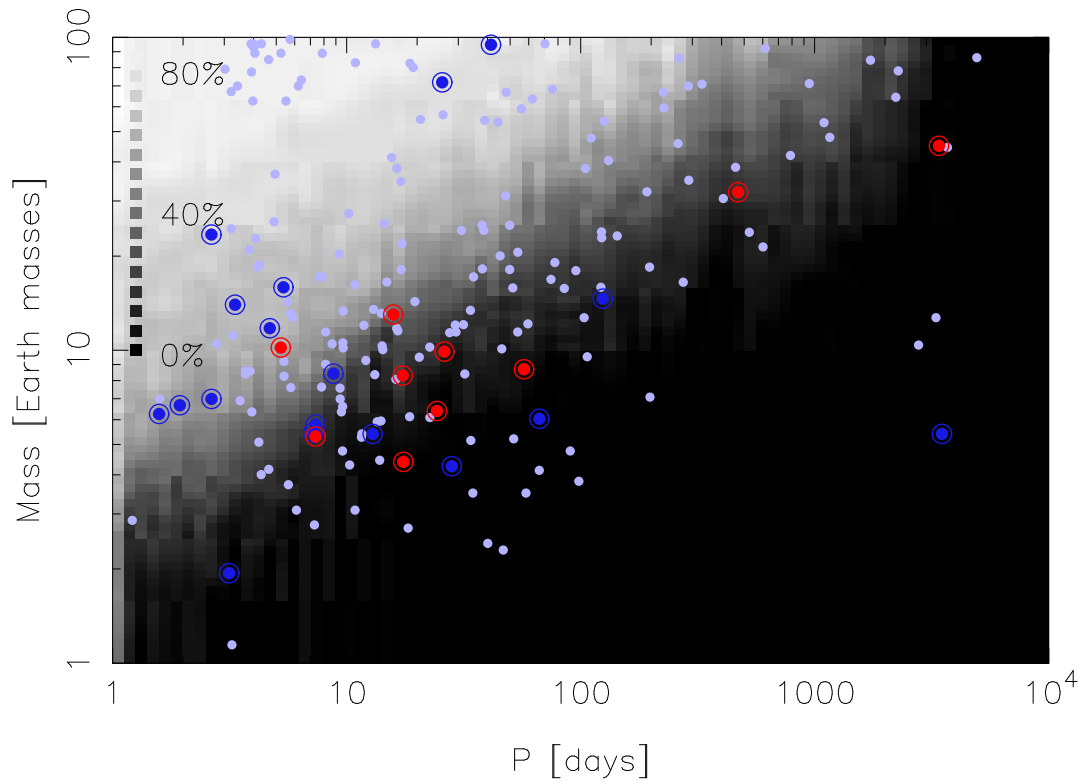}
\caption{Planet detection probability as determined by using Eq. (\ref{eq:detection}) in the combined UVES and HARPS data set as functions of orbital period and minimum mass. The various dots represent the known planets orbiting all stars (light blue dots), known planets orbiting M dwarfs (circled blue dots), and planet candidates in our sample (circled red dots). The detection probabilities do not exceed 85\% even at the high-mass short-period corner of the plot because there are six data sets where planetary signals could not be detected at all due a combination of low number of measurements and evidence of a massive companion that prevented detections of additional companions due to overparameterisation of the benchmark model.}\label{fig:sample_population}
\end{figure*}

We have plotted the known planets\footnote{Obtained from the Extrasolar Planets Encyclopaedia \citep[see][]{schneider2011}.} (light blue dots), known planets around M dwarfs (blue circled dots), the new candidates we report (red circled dots) in Fig. \ref{fig:sample_population}. This figure also shows the estimated detection probability of the combined UVES and HARPS data sets as functions of minimum mass and orbital period for the whole sample (Section \ref{sec:occurrence_computations}). The most remarkable feature in this plot appears to be that the new candidates are concentrated around minimum masses of 10 M$_{\oplus}$ and periods from few to few dozen days. Considering that such planets can be detected in this sample with rather low probabilities of roughly 10-30\%, the abundance of such companions appears strikingly high. Yet, the results from analyses of \emph{Kepler's} data appear to imply that this is a real feature in general \citep{howard2012}, and applies to M dwarfs in particular, as shown in \citet{dressing2013}. They observed an abundance of transiting planets with the same period range and radii of up to 3 R$_{\oplus}$, which could correspond to the same population of planets of which we only observe the planets with the highest masses.

The detection probability of a given combination of minimum mass and period in Fig. \ref{fig:sample_population} shows the different areas in the mass-period space where planets can be spotted easily and where it is very difficult (or unlikely) given the current sample. The highest gradient in this probability occurs along the line increasing from 4 M$_{\oplus}$ at a period of one day to 100 M$_{\oplus}$ at 1000 days. In our sample, only one candidate can be found above this line, even though planets in this region would be the easiest ones to detect. We note that the rather artificial-looking vertical line at roughly 3000 days in the top right corner of the Fig. \ref{fig:sample_population} is in fact a threshold arising from the fact that we limited the period search to periods at most the baselines of the data sets. Furthermore, there is a weakly distinguishable feature that shows decreased detection probabilities for periods around 365-day period that demonstrates how the existence of planetary signals at one year period cannot be ruled out generally as easily as slightly shorter and longer ones due to poor phase-coverage caused by data samplings.

Based on the analyses of the combined UVES and HARPS radial velocities of the 41 nearby M dwarfs, these data sets contain the signals of eight new exoplanet candidates out of which seven can be classified as super-Earths due to their minimum masses that are higher than that of the Earth but still lower, for most candidates considerably so, than one Neptune-mass. In addition to these super-Earths, we report a detection of a more massive sub-Saturnian companion orbiting GJ 229, and confirm the existence of the long-period planet orbiting GJ 443 detected by \citet{delfosse2012}. Eight of these candidate planets are thus new and previously unknown. We have listed the obtained orbital parameters together with the inferred minimum masses and semi-major axes of the new planet candidates we report in Table \ref{tab:planet_orbits}. The minimum masses and semi-major axes have been estimated by using the stellar masses as presented by \citet{zechmeister2009} and by assuming conservatively that their standard deviations are 10\% of the estimated values. The last column of Table \ref{tab:planet_orbits} shows the estimated locations of the planets, i.e. whether they are in the habitable zone \citep{kopparapu2013}, in the cool zone outside the outer edge of the habitable zone, or in the hot zone inside the inner edge of the habitable zone. We have also plotted the estimated log-posterior densities as functions of signal periods, indicative of significant periodicities, in the Appendix (Figs. \ref{fig:GJ27.1_solution} - \ref{fig:GJ682_solution}) together with phase-folded signals and probability distributions for the periods and signal amplitudes that demonstrate that the signals satisfy the detection criteria.

\begin{table*}
\caption{Orbital solutions (maximum \emph{a posteriori} estimates and the 99\% Bayesian credibility intervals) of the new planet candidates around M dwarfs and the inferred minimum masses and semi-major axes.}\label{tab:planet_orbits}
\begin{minipage}{\textwidth}
\begin{center}
\begin{tabular}{lcccccccc}
\hline \hline
Planet & $P$ & $K$ & $e$\footnote{The point and uncertainty estimates of the orbital eccentricities are dominated by the prior density of the eccentricity that favours circular orbits over eccentric ones.} & $\omega$\footnote{Bayesian credibility intervals are not tabulated for the angular parameters because they are equivalent to the defined ranges of these parameters of [0, 2$\pi$] for such close-circular orbits.} & $M_{0}$ & $m_{p} \sin i$ & $a$ & Zone\footnote{(HZ) habitable-zone planet according to the estimated inner and outer edges of \citep{kopparapu2013}; (C) cool planet, i.e. outside the outer edge of the HZ; (H) hot planet, i.e. inside the inner edge of the HZ.} \\
& [days] & [ms$^{-1}$] & & [rad] & [rad] & [M$_{\oplus}$] & [AU] \\
\hline
GJ 27.1 b & 15.819 [15.793, 15.842] & 4.90 [2.61, 6.30] & 0.08 [0, 0.26] & 2.7 & 5.8 & 13.0 [6.4, 17.1] & 0.101 [0.088, 0.110] & H \\
GJ 160.2 b & 5.2354 [5.2289, 5.2381] & 4.58 [1.99, 7.88] & 0.06 [0, 0.26] & 6.2 & 0.2 & 10.2 [4.3, 17.4] & 0.053 [0.046, 0.057] & H \\
GJ 180 b & 17.380 [17.360, 17.398] & 3.33 [1.46, 4.95] & 0.11 [0, 0.25] & 0.2 & 5.0 & 8.3 [3.0, 11.8] & 0.103 [0.089, 0.109] & H \\
GJ 180 c & 24.329 [24.263, 24.381] & 2.31 [0.92, 3.80] & 0.09 [0, 0.29] & 4.1 & 2.1 & 6.4 [2.3, 10.1] & 0.129 [0.112, 0.136] & HZ \\
GJ 229 b & 471 [459, 493] & 3.83 [2.15, 5.57] & 0.10 [0, 0.32] & 2.6 & 4.0 & 32 [16, 49] & 0.97 [0.88, 1.09] & C \\
GJ 422 b & 26.161 [26.063, 26.243] & 4.49 [2.70, 6.45] & 0.05 [0, 0.26] & 4.3 & 1.3 & 9.9 [5.9, 15.5] & 0.119 [0.108, 0.133] & HZ \\
GJ 433 b & 7.3697 [7.3661, 7.3731] & 2.60 [2.04, 3.59] & 0.05 [0, 0.21] & 5.4 & 3.6 & 5.3 [3.4, 7.3] & 0.060 [0.052, 0.064] & H \\
GJ 433 c\footnote{The orbit cannot be constrained from above but there is little doubt about the existence of this signal in the combined data, and it has been detected by using a larger set of HARPS data \citep{delfosse2012}, thus we list it as a candidate planet.} & 3400 [1900, --] & 2.90 [0.10, --] & 0.08 [0, 0.24] & 5.9 & 1.5 & 45 [1, --] & 3.6 [2.2, --] & C \\
GJ 682 b & 17.478 [17.438, 17.540] & 2.99 [1.14, 4.57] & 0.08 [0, 0.27] & 1.5 & 0.5 & 4.4 [2.0, 8.1] & 0.080 [0.076, 0.094] & HZ \\
GJ 682 c & 57.32 [56.84, 57.77] & 3.98 [1.82, 5.60] & 0.10 [0, 0.29] & 5.6 & 4.3 & 8.7 [4.1, 14.5] & 0.176 [0.167, 0.206] & C \\
\hline \hline
\end{tabular}
\end{center}
\end{minipage}
\end{table*}

We also broadly verified the existence of the signals corresponding to the planet candidates in Table \ref{tab:planet_orbits} by using an independent statistical method. We applied the log-likelihood periodograms of \citet{baluev2009} and \citet{anglada2013} and used the same statistical model as in the Bayesian analyses. The results of these periodogram analyses are shown in Table \ref{tab:likelihood_periodograms}. Six of the signals are clear with FAPs below a threshold of 0.1\%. However, GJ 180 c, GJ 422 b, and the two candidates of GJ 682 are only detected with FAPs greater than this threshold. While we suspect that the high FAP of GJ 422 b might be due to priors that were all assumed to be flat when calculating the log-likelihood periodograms and the assumptions behind the significance tests of these periodograms \citep{baluev2009}, it appears to be clear why GJ 180 c and the candidates GJ 682 b and c were not detected with the periodograms. This is because for GJ 180 c and GJ 682 c, the solution is actually obtained by assuming that there is only one signal when in reality there are two and the $k=1$ model is therefore fitted to the superposition of the two signals that causes an apparent decrease to the significance of the second signal. Instead, the signal of GJ 682 b could not even be detected (it was also below the detection threshold with the Bayesian tools because the log-Bayesian evidence ratio corresponding to the detection threshold of 10$^{4}$ is 9.21, see Table \ref{tab:UVES_signals}) because the $k=1$ model fitted the data much more poorly than a $k=2$ model. With the posterior samplings, however, detecting these signals was possible (Figs. \ref{fig:GJ180_solution} and \ref{fig:GJ682_solution}).

\begin{table}
\caption{Solutions obtained by using the log-likelihood periodograms. Preferred periods ($P$), false alarm probabilities (FAP), and the differences between the (natural) logarithms of the likelihoods of the preferred model with $k$ signals and a model with $k-1$ signals ($\Delta_{\rm L}$). \label{tab:likelihood_periodograms}}
\begin{minipage}{0.45\textwidth}
\begin{center}
\begin{tabular}{lccc}
\hline \hline
Planet & $P$ & FAP & $\Delta_{\rm L}$ \\
 & [days] & [\%] \\
\hline
GJ 27.1 b & 15.822 & 0.01 & 19.04 \\
GJ 160.2 b & 5.236 & 0.01 & 19.24 \\
GJ 180 b & 17.378 & 0.03 & 18.33 \\
GJ 180 c & 24.309 & 4.83 & 13.33 \\
GJ 229 b & 480.185 & 4.1$\times 10^{-4}$ & 23.12 \\
GJ 422 b & 26.143 & 1.4 & 14.97 \\
GJ 433 b & 7.372 & 9.4$\times 10^{-11}$ & 38.1 \\
GJ 433 c & 2947.6 & 5.2$\times 10^{-3}$ & 19.64 \\
GJ 682 b\footnote{The first signal in the GJ 682 data could not be detected because a $k=1$ model fits the combined data poorly due to a superposition of two signals with almost equal amplitudes.} & 56.96 & -- & -- \\
GJ 682 c & 17.465 & 1.7 & 14.59 \\
\hline \hline
\end{tabular}
\end{center}
\end{minipage}
\end{table}

\subsection{Potential additional signals}

The sample also contained ten additional signals that were well-constrained in period and amplitude but that did not exceed the detection threshold of $s = 10^{4}$ or were supported by data from only one instrument. We call these signals SRCs (signals requiring confirmation). They did, however, all exceed a less conservative threshold of $s = 150$ and it would thus be wrong to simply ignore the existence of these ''emerging'' signals. We took these signals into account when calculating the detection thresholds but we do not (yet) consider them to be candidate planets because we wish to avoid overestimation of the occurrence rates. However, we do calculate the occurrence rates under the assumption that these additional signals are planet candidates but remain cautious when interpreting the corresponding results as some of these signals might be false positives caused by noise, insufficient modelling, stellar activity, and/or instrumental artefacts.

The ten additional signals are found in the combined data of GJ 433 (at a period of 36.0 days), GJ 551 (332 and 2200 days), GJ 821 (12.6 days), GJ 842 (190 days), GJ 855 (12.7 and 26.2 days), GJ 1009 (24.5 days), GJ 1100 (34.4 days) and in the UVES data of GJ 891 (30.6 days).

\subsection{Occurrence rates}

When calculating the planet occurrence rates as described in Section \ref{sec:occurrence_computations}, we obtained some interesting estimates and show them in Table \ref{tab:occurrence}. We divided the period space into four bins: between 1 and 10 days, 10 and 100 days, 100 and 1000 days, and 1000 and 10$^{4}$ days and the minimum mass space into three bins between 3 and 10 $M_{\oplus}$, 10 and 30 $M_{\oplus}$, and 30 and 100 $M_{\oplus}$ (the last bin for masses above 100 M$_{\oplus}$ is omitted from the table). The resulting occurrence rates show several features that can be considered as representative of the underlying population of planets around M dwarfs. We also show the detection probabilities of each bin based on the whole sample in the bottom of Table \ref{tab:occurrence} for comparison.

\begin{table*}
\caption{Expected numbers of planets per star based on the sample of stars studied in the current work (top), potential numbers assuming that the ten signals requiring confirmation (SRCs) correspond to planet candidates (middle), and detection probability (DP) of the whole sample in each mass-period bin (bottom).\label{tab:occurrence}}
\begin{minipage}{\textwidth}
\begin{center}
\begin{tabular}{lcccc}
\hline \hline
Planet candidates & 1 day $< P \leq 10$ days & 10 days $< P \leq 100$ days & 100 days $< P \leq 1000$ days & 1000 days $< P$ \\
\hline
30 M$_{\oplus} < m_{p} \sin i \leq$ 100 M$_{\oplus}$ & 0.000$^{+0.048}_{-0.000}$ & 0.00$^{+0.17}_{-0.00}$ & 0.05$^{+0.06}_{-0.02}$ & 0.19$^{+0.19}_{-0.14}$ \\
10 M$_{\oplus} < m_{p} \sin i \leq$ 30 M$_{\oplus}$ & 0.037$^{+0.011}_{-0.006}$ & 0.06$^{+0.11}_{-0.03}$ & 0.00$^{+0.37}_{-0.00}$ & 0.0$^{+1.9}_{-0.0}$ \\
3 M$_{\oplus} < m_{p} \sin i \leq$ 10 M$_{\oplus}$ & 0.06$^{+0.11}_{-0.03}$ & 1.02$^{+1.48}_{-0.69}$ & 0.00$^{+0.42}_{-0.00}$ & 0.0$^{+2.0}_{-0.0}$ \\
\hline
Planet candidates and SRCs\footnote{Values omitted from bins without candidates or SRCs.} \\
\hline
30 M$_{\oplus} < m_{p} \sin i \leq$ 100 M$_{\oplus}$ & -- & 0.036$^{+0.014}_{-0.007}$ & 0.11$^{+0.11}_{-0.05}$ & 0.19$^{+0.19}_{-0.15}$ \\
10 M$_{\oplus} < m_{p} \sin i \leq$ 30 M$_{\oplus}$ & 0.037$^{+0.011}_{-0.006}$ & 0.30$^{+0.53}_{-0.14}$ & -- & -- \\
3 M$_{\oplus} < m_{p} \sin i \leq$ 10 M$_{\oplus}$ & 0.06$^{+0.11}_{-0.03}$ & 1.43$^{+2.07}_{-0.96}$ & 0.35$^{+0.15}_{-0.19}$ & 1.8$^{+1.8}_{-1.5}$ \\
\hline
Sample DP \\
\hline
30 M$_{\oplus} < m_{p} \sin i \leq$ 100 M$_{\oplus}$ & 0.74 & 0.69 & 0.46 & 0.13 \\
10 M$_{\oplus} < m_{p} \sin i \leq$ 30 M$_{\oplus}$ & 0.65 & 0.40 & 0.17 & 0.04 \\
3 M$_{\oplus} < m_{p} \sin i \leq$ 10 M$_{\oplus}$ & 0.40 & 0.12 & 0.07 & 0.01 \\
\hline \hline
\end{tabular}
\end{center}
\end{minipage}
\end{table*}

According to the estimated occurrence rates in Table \ref{tab:occurrence} (top), the occurrence rate of super-Earths and more massive planets with $m_{p} \sin i$ of up to 30 M$_{\oplus}$ increases dramatically for periods between 10 and 100 days. We find that the occurrence rate of planets with masses between 10 and 30 M$_{\oplus}$ is 0.06$^{+0.11}_{-0.03}$ for the sample in this period range. The most dramatic occurrence rate can be found for super-Earths (3 M$_{\oplus} < m_{p} \sin i \leq 10$ M$_{\oplus}$) on orbits between 10 and 100 days of 1.02$^{+1.48}_{-0.69}$ per star, which indicates that such planets are very common around M dwarfs in the Solar neighbourhood. Comparison with the results of \citet{bonfils2013}, which is a similar survey of a larger sample of nearby M dwarfs in the southern sky, is difficult because they did not detect any candidates with masses between 10 and 100 M$_{\oplus}$ and periods in excess of 10 days. They did, however, detect two candidates with masses between 1 and 10 M$_{\oplus}$ and periods between 10 and 100 days, which implied an occurrence rate of such companions of 0.54$^{+0.50}_{-0.16}$. These estimates are broadly consistent when bearing in mind the different sensitivities of the approaches and the fact that we were able to detect more such companions in the sample than \citet{bonfils2005} could in their larger sample of 104 M dwarfs.

Second, the occurrence rate of planets more massive than 10 M$_{\oplus}$ on orbits with periods less than 10 days is very low, roughly 0.037$^{+0.011}_{-0.006}$ planets per star for planets less massive than 30 M$_{\oplus}$. In comparison to the results obtained by the HARPS M dwarf survey, \citet{bonfils2013} reported the the occurrence rate of planets with periods from 1 to 10 days and masses between 10 and 100 M$_{\oplus}$ to be 0.03$^{+0.04}_{-0.01}$, which is consistent with our estimate of 0.037$^{+0.011}_{-0.006}$. This result is also consistent with the observation of \citet{dressing2013} that the occurrence rate of planets with radii in excess of 2.0 R$_{\oplus}$ on such orbital periods is roughly 0.05 and is therefore very likely a general feature for M dwarfs. However, comparing our results and the results of \citet{bonfils2013} with \emph{Kepler} occurrence rates based on estimated planetary radii cannot be performed confidently without bias. We find only a slighly higher occurrence rate of 0.06$^{+0.11}_{-0.03}$ for planets with masses below 10 M$_{\oplus}$ because there was only one such candidate in our sample.

These results are very difficult to compare with the results of \citet{dressing2013} and \citet{morton2013} because the mass-radius relation is far from a well-established one for a range of masses and radii in the super-Earth regime. Yet, \citet{dressing2013} found the occurrence rate of planets to increase by a factor of ten when moving from radii interval of 2.8 - 5.7 R$_{\oplus}$ down to 1.4 - 2.8 R$_{\oplus}$, which is as dramatic increase as we find when moving from masses between 10 and 30 M$_{\oplus}$ down to the interval between 3 and 10 M$_{\oplus}$. While this does not mean that the results are consistent, it suggests that this large change in abundance may have the same origin. Obviously, this depends on whether the low-temperature \emph{Kepler} sample and the M dwarf sample analysed in the current work are drawn from the same population. Unlike the \emph{Kepler} sample which comprises of more massive and brighter M dwarfs further out from the galactic plane, samples analysed in the current work and in \citet{bonfils2013} are likely to be approximately volume-limited.

Third, our results suggest that planets with masses below 100 M$_{\oplus}$ and periods longer than 100 days might be abundant around nearby M dwarfs. However, conclusions are very hard to draw at the moment because of the low number of such candidates in the combined UVES and HARPS data. \citet{bonfils2013} did not observe any such planets in their sample even though the candidate orbiting GJ 433 could have been detected had they combined the HARPS data with the UVES velocities that were published in 2009. The same team of researchers reported this candidate in \citet{delfosse2012}, which means it could have been included in the results of \citet{bonfils2013}.

Finally, out of the ten planet candidates in the sample, three appear to have orbits located within the respective stellar habitable zones according to the equations of \citet{kopparapu2013} for the inner (moist greenhouse) and outer (maximum greenhouse) edge of the habitable zone. This enables us to state that 30\% of M dwarf planets in the current sample are located within the HZs. However, we have to take into account the detection probability of the current data sets of habitable-zone planet candidates to estimate the occurrence rate of habitable-zone planets in a statistically representative, and thus meaningful, manner. We achieve this by calculating the detection probabilities in Fig. \ref{fig:sample_population} as a function of semi-major axis instead of orbital period for each star and combine the resulting probabilities in the same way as in Section \ref{sec:occurrence_computations} but by limiting the analysis to the stellar habitable zones listed in Table \ref{tab:target_properties}. We find that the occurrence rate of planets with minimum masses between 3 and 10 M$_{\oplus}$ in the habitable zones of the sample stars is 0.21$^{+0.03}_{-0.05}$ planets per star and that of planets with minimum masses between 10 and 30 M$_{\oplus}$ is 0.035$^{+0.013}_{-0.007}$ planets per star.

These estimates can be compared with the results of \citet{bonfils2013} according to whom M dwarfs would have on average 0.41$^{+0.54}_{-0.13}$ super-Earths (1 M$_{\oplus} < m_{p} \sin i < 10$ M$_{\oplus}$) per star in the habitable zones, which appears to be consistent with our estimate considering the slightly different mass range. These estimates also appear to be an order of magnitude greater than the occurrence rate estimates from the low-temperature sample of \emph{Kepler} stars \citep{dressing2013}. In particular, \citet{dressing2013} estimated that the occurrence rate of Earth-size planets (0.5 - 1.4 R$_{\oplus}$) is 0.06$^{+0.06}_{-0.03}$ and that of larger ones (1.4 - 4 R$_{\oplus}$) is 0.03$^{+0.05}_{0.02}$ in the habitable zones of such stars. However, \citet{kopparapu2013b} revised the estimates of \citet{dressing2013} by calculating the habitable zones according to the modified equations of \citet{kopparapu2013}. As a result, there are 0.48$^{+0.12}_{-0.24}$ planets with 0.5 R$_{\oplus} < r_{p} <$ 1.4 R$_{\oplus}$ per star in the habitable zones of the stars in the sample of \citet{dressing2013}, which increases to 0.51$^{+0.10}_{-0.20}$ when increasing the radius range to 2 R$_{\oplus}$. These estimates are broadly consistent with our results.

\begin{figure}
\center
\includegraphics[width=0.45\textwidth]{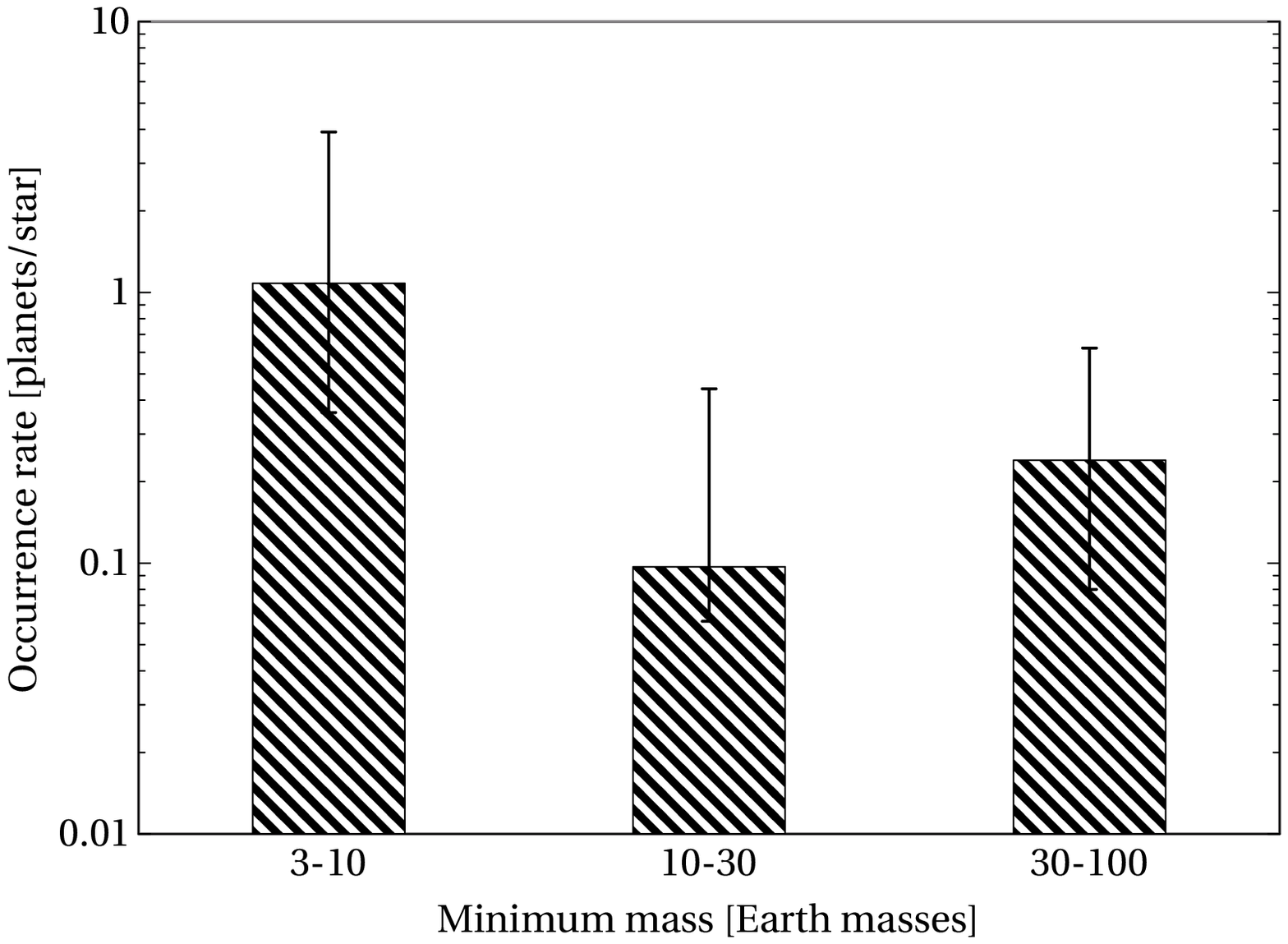}
\includegraphics[width=0.45\textwidth]{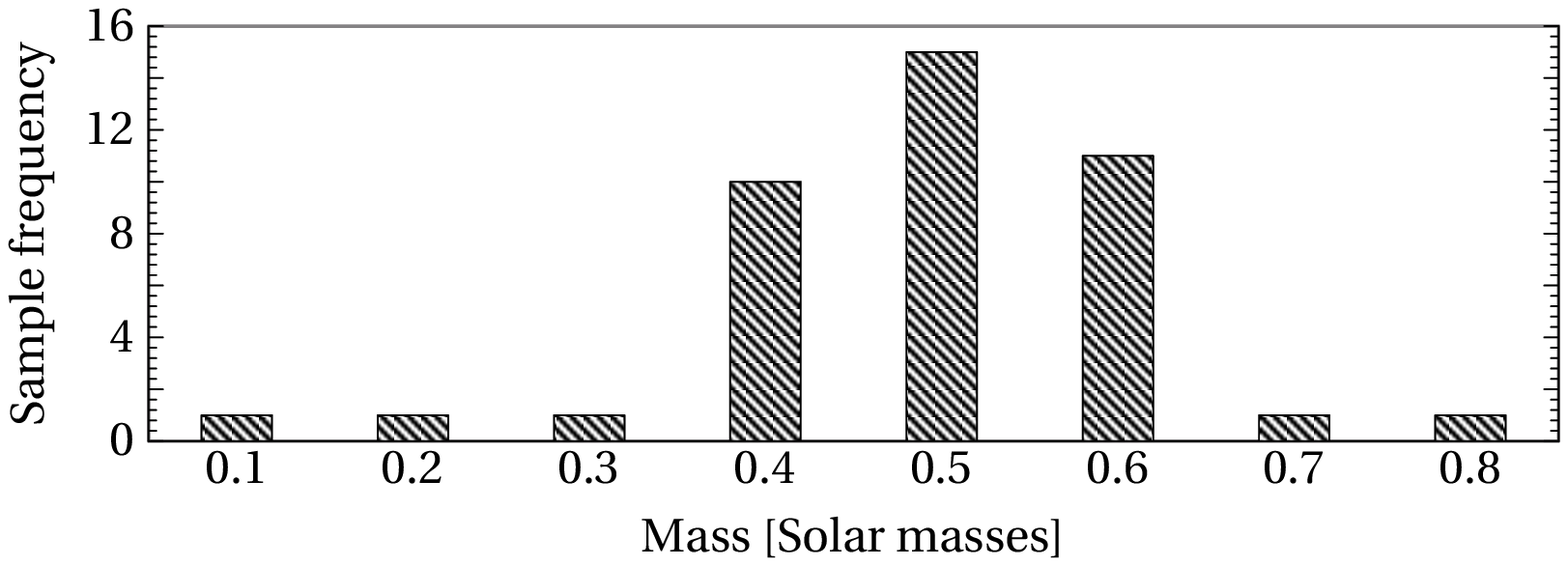}
\caption{Estimated mass-function of low-mass planets (top) based on the planet candidates in the sample and the mass-distribution of the M dwarfs in the sample (bottom).}\label{fig:mass_function}
\end{figure}

We also estimated the planetary mass function for low-mass companions with $m_{p} \sin i < 100$ M$_{\oplus}$. We used the same minimum mass bins as in Table \ref{tab:occurrence} and plotted the resulting occurrence rate as a function of minimum mass in Fig. \ref{fig:mass_function} (top panel) together with the mass distribution of the stars in the sample (bottom panel). This figure shows that the mass function increases dramatically with decreasing mass, similar to the increase found for planets orbiting more massive stars \citep{lopez2012}, which suggests that the high occurrence rate of super-Earths with $m_{p} \sin i <$ 10 M$_{\oplus}$ corresponds to the rapid increase observed in the \emph{Kepler} transit data for planets with radii less than 4 R$_{\oplus}$ \citep{morton2013}. However, the uncertainties are still too high to quantify this increase in a meaningful way and therefore we do not attempt to estimate the mass function quantitatively. Our occurrence rates denoted using the shaded histogram in Fig. \ref{fig:mass_function} (top panel) for minimum masses between 3 - 10, 10 - 30, and 30 -100 M$_{\oplus}$ are 1.08$^{+2.83}_{-0.72}$, 0.10$^{+0.35}_{-0.04}$, and 0.24$^{+0.39}_{0.16}$ planets per star, respectively.

When calculating the occurrence rates under the assumption that the additional ten emerging signals, or SRCs, in the sample are also caused by planets, we obtained even higher occurrence rates for the low-mass planets around M dwarfs (Table \ref{tab:occurrence}, middle). These numbers are consistent with but slightly higher than those obtained for the ten candidate planets in the sample. However, we note that some of these signals could be false positives.

\section{New planetary systems}\label{sec:new_systems}

We have listed the new planet candidates from our analyses of the sample velocities in Table \ref{tab:planet_orbits}. In this section, we discuss them briefly.

\subsection{GJ 433}

According to our results, we could also identify the two planetary signals that have been reported in the HARPS data of GJ 433 \citep{delfosse2012}. We did not, however, detect any additional candidate planets satisfying our detection criteria in the combined HARPS and UVES data of the star.

We note that \citet{bonfils2013} also reported signals in the HARPS data around 30 days although there are significant problems with their tabulated solutions\footnote{In their Table 7, \citet{bonfils2013} listed solutions that (i) have velocity amplitude estimates 100 times lower than their uncertainties, e.g. $K = $ 31.6 $\pm$ 3800.5 ms$^{-1}$ for Gl 54.1, implying that these cannot be real solutions in any meaningful way, and (ii) eccentricity estimates close to or equal to unity -- the latter being impossible for periodicities -- with occasional uncertainty estimates in excess of unity, e.g. $e = 0.9 \pm 8.9$ for Gl 54.1.}. Nevertheless, the log-posterior of this target shows an emerging maximum at a period of 36 days for a three-Keplerian model together with a local maximum at 50 days, which suggests that a two-Keplerian model might not be a sufficiently accurate description of the combined UVES and HARPS velocities. If either one (or both) of these emerging signals is confirmed by future data, GJ 433 would become one of the highly populated planetary systems around M dwarfs together with the famous planet hosting stars GJ 581, GJ 667C, GJ 163, and GJ 676A.

\subsection{GJ 682}

In our sample, there are also two stars with two new candidate super-Earths orbiting them in the period space between 10 and 100 days (Table \ref{tab:planet_orbits}). GJ 682 is orbited by candidates with minimum masses of 4.4 [2.0, 8.1] and 8.7 [4.1, 14.5] M$_{\oplus}$ on orbits with periods of 17.478 [17.438, 17.540] and 57.32 [56.84, 57.77] days, respectively. The former candidate is located in the stellar habitable zone and can be classified as a habitable-zone super-Earth, although its minimum mass is consistent with (sub) Neptunian structure. An interesting detail in the model probabilities of the GJ 682 velocities is that the one-Keplerian model is only 9900 times more probable than the model without any signals, which implies that a one-Keplerian model is not sufficiently good description of the data to enable the detection of a planet candidate according to our criteria. However, a two-Keplerian model has a 1.8$\times 10^{7}$ times greater probability than the one-Keplerian model, which implies that there is very strong evidence in favour of two candidates orbiting the star. This means that when there are at least two signals of similar amplitude in the velocities, models with only one signal can be difficult to use to interpret the data because they are not good enough in describing the velocity variations.

GJ 682 is an inactive dwarf star with no signs of chromospheric activity \citep{walkowicz2009} and has a projected rotation period of 10.7 days \citep{reiners2007}. This suggests that the signals we have detected indeed are Doppler signatures of planets.

We note that possible additional signals in the GJ 682 data, and in particular their superpositions with the two detected ones, might cause biases to the obtained parameters of the two candidates listed in Table \ref{tab:planet_orbits}.

\subsection{GJ 180}

Another system with two candidate planets, GJ 180, corresponds to a remarkable configuration of super-Earths with an orbital period ratio of 7:5 -- orbital periods of 17.380 [17.360, 17.398] and 24.329 [24.263, 24.381] days, respectively. This suggests (although does not imply) the existence of a stabilising 7:5 mean motion resonance \citep[see also][]{jenkins2013a}. The outer candidate is located in the stellar habitable zone, which makes this candidate interesting because of its reasonably low minimum mass of 6.4 [2.3, 10.1] M$_{\oplus}$ that enables its classification as a habitable-zone super-Earth. This system, its detection, dynamical stability, and formation history, are discussed in detail in a separate publication \citep{jenkins2013b}.

\subsection{GJ 422}

The candidate planet GJ 422 b has a minimum mass of 9.9 [5.9, 15.5] M$_{\oplus}$ and is thus a super-Earth or a sub-Neptunian planet, depending on the composition and atmospheric properties. Its orbit is likely located within the stellar habitable zone of GJ 422.

\subsection{GJ 27.1}

We also find a similar candidate orbiting GJ 27.1 with a minimum mass of 13.0 [6.4, 17.1] M$_{\oplus}$. This candidate can be classified as a sub-Neptunian planet candidate because it is likely too massive to be considered a super-Earth with rocky composition. With an orbital period of 15.819 [15.793, 15.842] days this signal is unlikely to be caused by stellar activity coupled with the rotation period whose projected estimate is 11.9 days \citep{houdebine2011}.

\subsection{GJ 160.2}

In addition to GJ 433 b, GJ 160.2 b is the only candidate in our sample with an orbital period shorter than 10 days. Candidates of this kind are the easiest ones to observe in our sample (see Fig. \ref{fig:sample_population}) and the fact that we only found two of them demonstrates that such planets are not very common around M dwarfs as also quantified by the occurrence rates in Table \ref{tab:occurrence}. We note that GJ 160.2 might actually be a K dwarf \citep{koen2010}, although \citet{zechmeister2009} classified it as a M0 V star. The candidate GJ 160.2 b is a hot sub-Neptunian planet.

The star has a projected rotation period of 43.6 days \citep{houdebine2011}, which indicates that the signal in unlikely to be related to stellar rotation and thus activity.

\subsection{GJ 229}

Finally, we report a discovery of a planet candidate orbiting GJ 229 with an orbital period of 471 [459, 493] days and a minimum mass of 32 [16, 49] M$_{\oplus}$. This discovery makes the GJ 229 system one of the most diverse systems around M dwarfs because \citet{nakajima1995} reported a brown dwarf companion to the star based on direct imaging.

\section{Conclusions}\label{sec:discussion}

We have presented our analysis of UVES velocities of a sample of 41 M dwarfs \citep{zechmeister2009} when combining the velocities with HARPS precision data as obtained from the spectra available in the ESO archive. As a result, we report the existence of eight new planet candidates around the sample stars (Tables \ref{tab:UVES_signals} and \ref{tab:planet_orbits}) and confirm the existence of the two companions around GJ 433 \citep{delfosse2012} that exceed our conservative probabilistic detection threshold by making the statistical models more than 10$^{4}$ times more probable than models without the corresponding signals. Among the most interesting targets in our sample are GJ 433, GJ 180, and GJ 682, with at least two candidate planets each.

We have also presented estimates for the occurrence rate of low-mass planets around M dwarfs (Table \ref{tab:occurrence}) based on the current sample. We find that low-mass planets are very common around M dwarfs in the Solar neighbourhood and that the occurrence rate of planets with masses between 3 and 10 M$_{\oplus}$ is 1.08$^{+2.83}_{-0.72}$ per star. This estimate is likely consistent with that suggested based on the \emph{Kepler} results for a sample of stars with $T_{\rm eff} < 4000$ K \citep{dressing2013,morton2013}, although the comparisons are not easily performed because we could assess the occurrence rates of companions with periods up to the span of the radial velocity data of a few thousand days. On the other hand, we confirm the lack of planets with masses above 3 M$_{\oplus}$ on orbits with periods between 1-10 days. Such companions to low-mass stars have an occurrence rate of only 0.06$^{+0.11}_{-0.03}$ planets per star based on our sample.

There are nine targets in the sample that are also found in the sample of M dwarfs presented in \citet{bonfils2013}: GJ 1, GJ 176, GJ 229, GJ 357, GJ 433, GJ 551, GJ 682, GJ 699, GJ 846, and GJ 849. Out of these nine stars, we found signals in the velocities of GJ 229, GJ 433, and GJ 682. Our results are essentially similar for GJ 433, for which \citet{bonfils2013} reported a signal at 7.4 days and the same group reported another long-period signal when analysing the HARPS data in combination with the UVES data analysed here \citep{delfosse2012}.

The planet candidates GJ 229 b, GJ 682 b and c have orbital periods of 471 [459, 493], 17.478 [17.438, 17.540], and 57.32 [56.84, 57.77] days. \citet{bonfils2013} did not report any such periodicities for these stars. We believe the reason is that we obtained HARPS-TERRA velocities from the HARPS spectra that are more precise for M dwarfs \citep{anglada2012c}, combined the HARPS velocities with the UVES ones which provides more information on the underlying periodic signals regardless of whether the signals can be detected in the two data sets independently or not, and accounted for correlations in the velocity data that could disable the detections of low-amplitude signals if not accounted for \citep{baluev2012,tuomi2012c,tuomi2013b}.

We have compared our results briefly with those obtained by using the \emph{Kepler} space-telescope \citep[e.g.][]{howard2012,dressing2013} in Section \ref{sec:planet_statistics}. However, such a comparison is not necessarily reliable because the properties of \emph{Kepler's} transiting planet candidates can only be discussed in terms of planetary radii and the radial velocity method can only be used to obtain minimum masses. Because of this, it is not surprising that there are remarkable differences that are unlikely to arise by chance alone. For instance, \citet{dressing2013} estimated that there are roughly 0.15$^{+0.13}_{0.06}$ Earth-sized planets (radii between 0.5 and 1.4 R$_{\oplus}$) in the habitable zones of cool stars (with $T_{\rm eff} <$ 4000 K) and that the nearest such planet could be expected to be found within 5 pc with 95\% confidence. We calculated a similar estimate for candidates with masses between 3 and 10 M$_{\oplus}$ and obtained an occurrence rate estimate of 0.21$^{+0.03}_{-0.05}$ planets per star that appears to be higher than the estimate of \citet{dressing2013} despite the fact that we cannot assess the occurrence rates of planets with masses below 3 M$_{\oplus}$ because we did not detect any such candidates orbiting the stars in the sample. However, these estimates can only be compared in detail with a range of robust planet composition and evolution models in hand, and is beyond the current work.

According to our results, M dwarfs have very high rates of hosting systems of low-mass planets around them and have a high probability of being hosts to super-Earths in their habitable zones. Together with the fact that radial velocity surveys can be used to obtain evidence for Earth-mass planets orbiting such stars, and the fact that M dwarfs are very abundant in the Solar neighbourhood, this makes them primary targets for searches of Earth-like planets, and possibly life, with current and future planet surveys.

\section*{Acknowledgements}

The authors acknowledge M. Zechmeister et al. for making the UVES velocity data of the sample stars available and the significant efforts of the HARPS-ESO team in improving the instrument and its data reduction pipelines. We also acknowledge the efforts of all the individuals that have been involved in observing the target stars with HARPS and UVES spectrographs because without such efforts, the current work would not have been possible. JSJ acknowledges funding by Fondecyt through grant 3110004 and partial support from CATA (PB06, Conicyt), the  GEMINI-CONICYT FUND and from the Comit\'e Mixto ESO-GOBIERNO DE CHILE.


\clearpage

\newpage

\appendix

\section{Orbital solutions}

\begin{figure*}
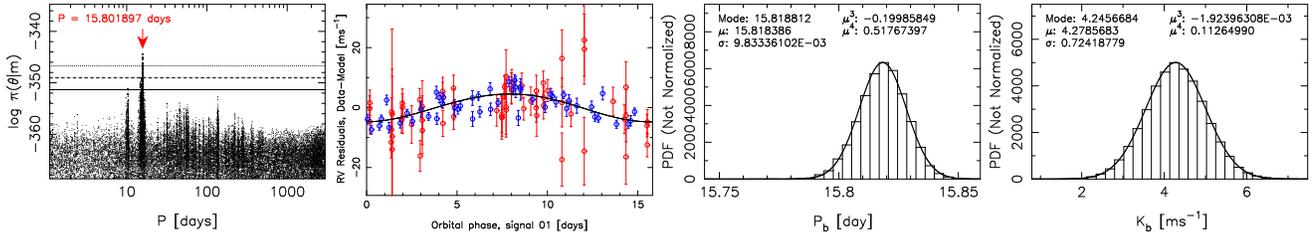

\center
\includegraphics[angle=270, width=0.24\textwidth]{rvdist01_GJ27.1b_psearch_b.ps}
\includegraphics[angle=270, width=0.24\textwidth]{rvdist01_scresidc_GJ27.1b_1.ps}
\includegraphics[angle=270, width=0.24\textwidth]{rvdist01_GJ27.1b_dist_Pb.ps}
\includegraphics[angle=270, width=0.24\textwidth]{rvdist01_GJ27.1b_dist_Kb.ps}
\caption{Solution for GJ 27.1 b. From left to right: estimated posterior density of the period from samplings with $\beta < 1$, where the horisontal lines denote probability thresholds of 10\% (dotted), 1\% (dashed), and 0.1\% (solid) of the MAP estimate denoted by the red arrow; phase-folded MAP radial velocity curve, where UVES and HARPS measurements are denoted using red and blue circles, respectively; and the marginalised posterior densities of the period and velocity amplitude, where the solid curves denote Gaussian functions with the same mean and variance.}\label{fig:GJ27.1_solution}
\end{figure*}

\begin{figure*}
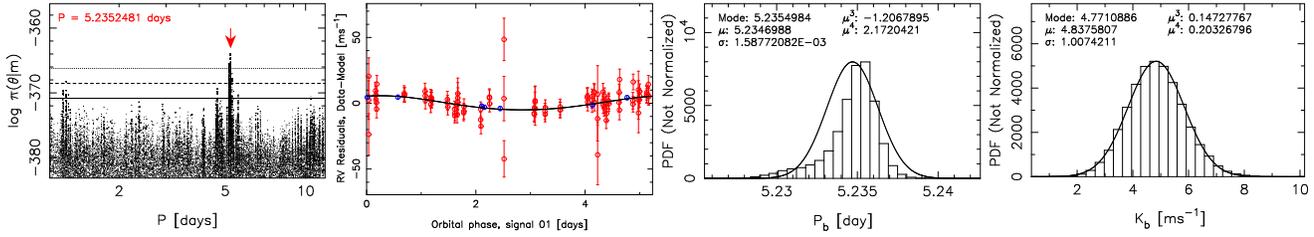

\center
\includegraphics[angle=270, width=0.24\textwidth]{rvdist01_GJ160.2b_psearch_b.ps}
\includegraphics[angle=270, width=0.24\textwidth]{rvdist01_scresidc_GJ160.2b_1.ps}
\includegraphics[angle=270, width=0.24\textwidth]{rvdist01_GJ160.2b_dist_Pb.ps}
\includegraphics[angle=270, width=0.24\textwidth]{rvdist01_GJ160.2b_dist_Kb.ps}
\caption{As in Fig. \ref{fig:GJ27.1_solution} but for GJ 160.2 b.}\label{fig:GJ160.2_solution}
\end{figure*}

\begin{figure*}
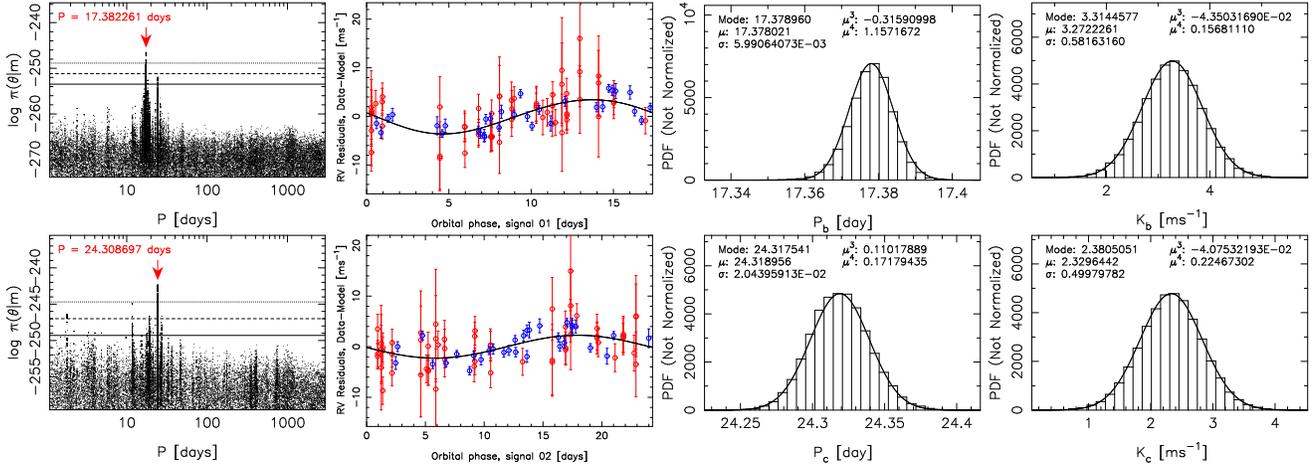

\center
\includegraphics[angle=270, width=0.24\textwidth]{rvdist01_GJ180b_psearch_b.ps}
\includegraphics[angle=270, width=0.24\textwidth]{rvdist02_scresidc_GJ180b_1.ps}
\includegraphics[angle=270, width=0.24\textwidth]{rvdist02_GJ180b_dist_Pb.ps}
\includegraphics[angle=270, width=0.24\textwidth]{rvdist02_GJ180b_dist_Kb.ps}

\includegraphics[angle=270, width=0.24\textwidth]{rvdist02_GJ180b_psearch_c.ps}
\includegraphics[angle=270, width=0.24\textwidth]{rvdist02_scresidc_GJ180b_2.ps}
\includegraphics[angle=270, width=0.24\textwidth]{rvdist02_GJ180b_dist_Pc.ps}
\includegraphics[angle=270, width=0.24\textwidth]{rvdist02_GJ180b_dist_Kc.ps}
\caption{As in Fig. \ref{fig:GJ27.1_solution} but for GJ 180 b (top) and c (bottom).}\label{fig:GJ180_solution}
\end{figure*}

\begin{figure*}
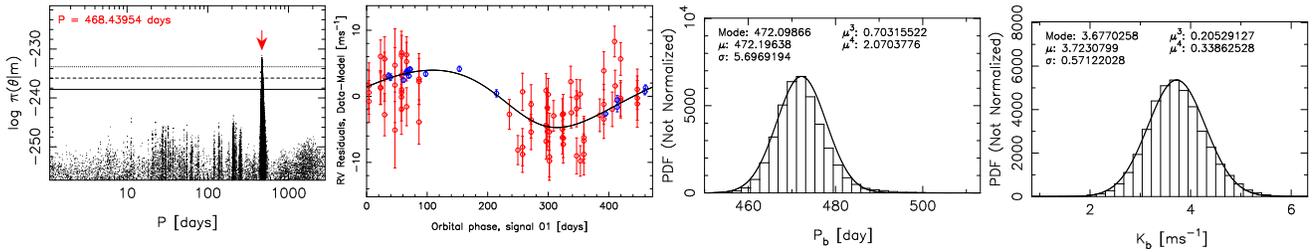

\center
\includegraphics[angle=270, width=0.24\textwidth]{rvdist01_GJ229b_psearch_b.ps}
\includegraphics[angle=270, width=0.24\textwidth]{rvdist01_scresidc_GJ229b_1.ps}
\includegraphics[angle=270, width=0.24\textwidth]{rvdist01_GJ229b_dist_Pb.ps}
\includegraphics[angle=270, width=0.24\textwidth]{rvdist01_GJ229b_dist_Kb.ps}
\caption{As in Fig. \ref{fig:GJ27.1_solution} but for GJ 229 b.}\label{fig:GJ229_solution}
\end{figure*}

\begin{figure*}
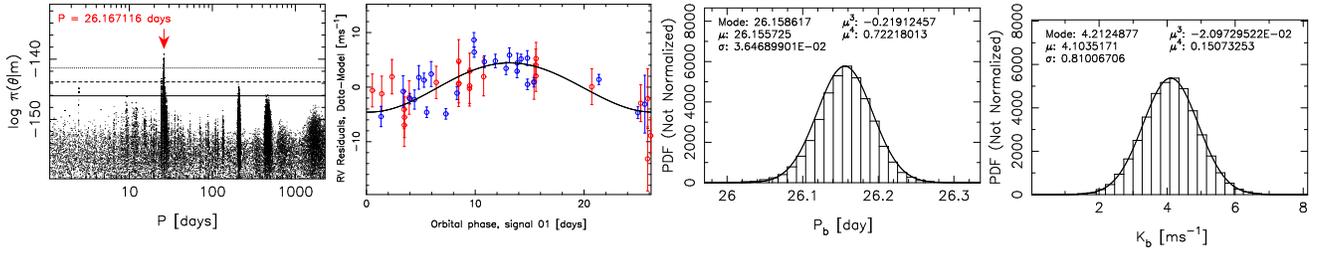

\center
\includegraphics[angle=270, width=0.24\textwidth]{rvdist01_GJ422b_psearch_b.ps}
\includegraphics[angle=270, width=0.24\textwidth]{rvdist01_scresidc_GJ422b_1.ps}
\includegraphics[angle=270, width=0.24\textwidth]{rvdist01_GJ422b_dist_Pb.ps}
\includegraphics[angle=270, width=0.24\textwidth]{rvdist01_GJ422b_dist_Kb.ps}
\caption{As in Fig. \ref{fig:GJ27.1_solution} but for GJ 422 b.}\label{fig:GJ422_solution}
\end{figure*}

\begin{figure*}
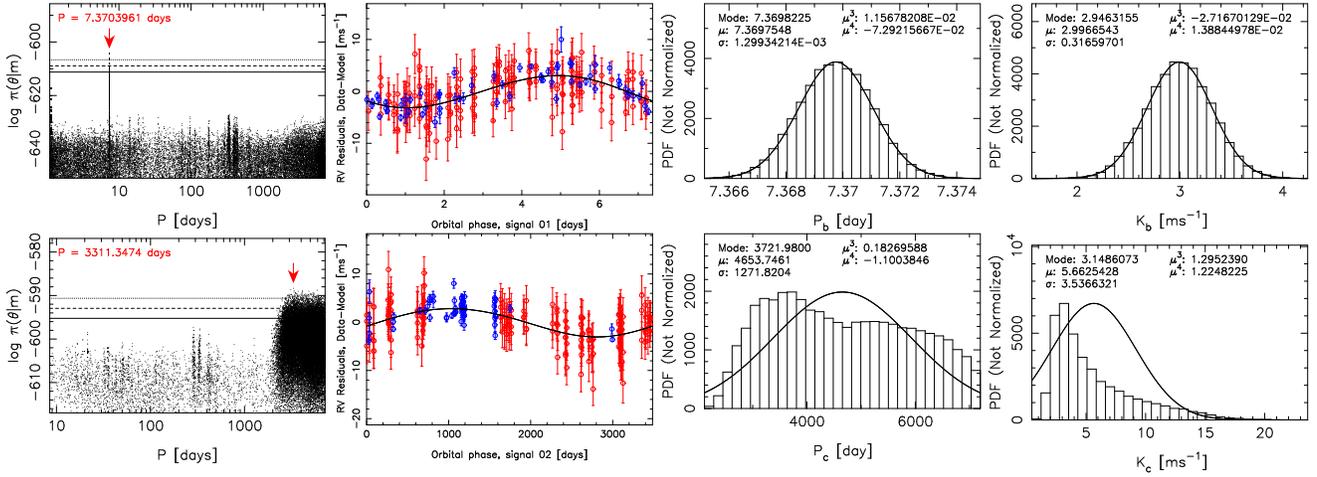

\center
\includegraphics[angle=270, width=0.24\textwidth]{rvdist01_GJ433b_psearch_b.ps}
\includegraphics[angle=270, width=0.24\textwidth]{rvdist02_scresidc_GJ433b_1.ps}
\includegraphics[angle=270, width=0.24\textwidth]{rvdist02_GJ433b_dist_Pb.ps}
\includegraphics[angle=270, width=0.24\textwidth]{rvdist02_GJ433b_dist_Kb.ps}

\includegraphics[angle=270, width=0.24\textwidth]{rvdist02_GJ433b_psearch_c.ps}
\includegraphics[angle=270, width=0.24\textwidth]{rvdist02_scresidc_GJ433b_2.ps}
\includegraphics[angle=270, width=0.24\textwidth]{rvdist02_GJ433b_dist_Pc.ps}
\includegraphics[angle=270, width=0.24\textwidth]{rvdist02_GJ433b_dist_Kc.ps}

\caption{As in Fig. \ref{fig:GJ27.1_solution} but for GJ 433 b (top) and c (bottom).}\label{fig:GJ433_solution}
\end{figure*}

\begin{figure*}
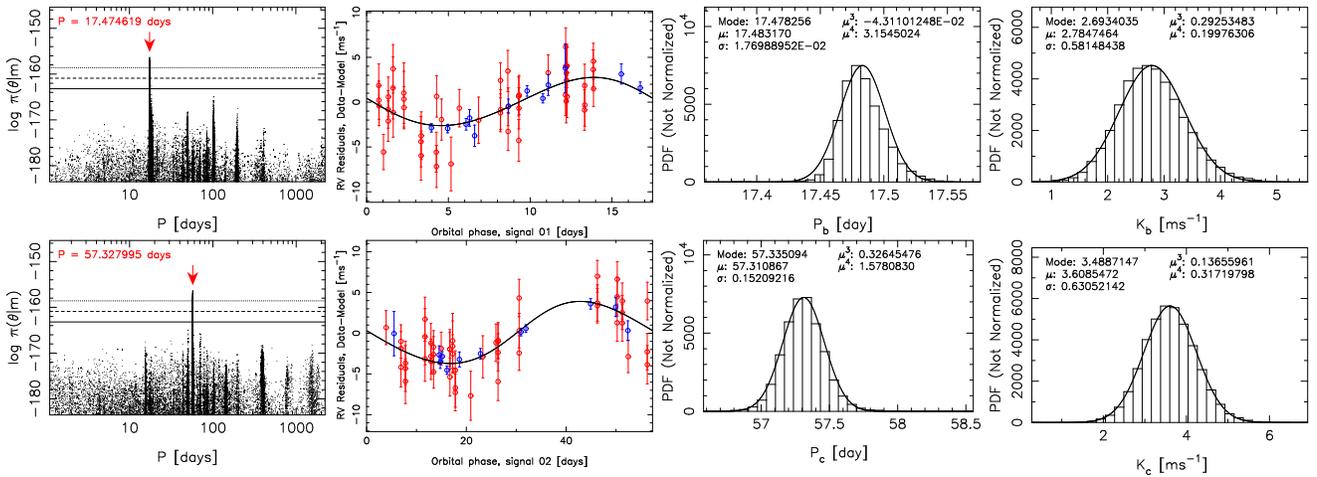

\center
\includegraphics[angle=270, width=0.24\textwidth]{rvdist02_GJ682b_psearch_b.ps}
\includegraphics[angle=270, width=0.24\textwidth]{rvdist02_scresidc_GJ682b_1.ps}
\includegraphics[angle=270, width=0.24\textwidth]{rvdist02_GJ682b_dist_Pb.ps}
\includegraphics[angle=270, width=0.24\textwidth]{rvdist02_GJ682b_dist_Kb.ps}

\includegraphics[angle=270, width=0.24\textwidth]{rvdist02_GJ682b_psearch_c.ps}
\includegraphics[angle=270, width=0.24\textwidth]{rvdist02_scresidc_GJ682b_2.ps}
\includegraphics[angle=270, width=0.24\textwidth]{rvdist02_GJ682b_dist_Pc.ps}
\includegraphics[angle=270, width=0.24\textwidth]{rvdist02_GJ682b_dist_Kc.ps}
\caption{As in Fig. \ref{fig:GJ27.1_solution} but for GJ 682 b (top) and c (bottom).}\label{fig:GJ682_solution}
\end{figure*}

\clearpage

\section{HARPS-TERRA velocities}\label{sec:velocities}

The HARPS-TERRA velocities obtained from the publicly available spectra in the ESO archive are presented in this section for all the targets for which at least two spectra were available. The secular acceleration has been subtracted from every HARPS-TERRA data set.

\begin{table}
\caption{HARPS-TERRA velocity data of GJ 1.}\label{tab:gj1_terra}
\begin{center}
\begin{tabular}{lcc}
\hline \hline
Time & Velocity & Unc. \\

[JD-2400000] & [ms$^{-1}$] & [ms$^{-1}$] \\
\hline
52985.5958 & -2.00 & 0.77 \\
52998.5772 & -4.37 & 0.17 \\
53206.8961 & -3.01 & 0.42 \\
53335.6181 & -0.48 & 0.62 \\
53520.9338 & 1.67 & 1.64 \\
53572.9309 & -1.50 & 0.46 \\
53575.8708 & -1.86 & 0.35 \\
53668.6705 & 2.40 & 0.42 \\
53672.6842 & -0.42 & 0.37 \\
53692.6006 & -0.04 & 0.65 \\
53694.6156 & -0.90 & 0.62 \\
53700.5993 & -1.43 & 0.50 \\
53721.5636 & 2.93 & 0.86 \\
54291.9230 & -3.84 & 0.56 \\
54295.8909 & -3.30 & 0.43 \\
54341.8116 & 0.89 & 0.54 \\
54343.8380 & 1.76 & 0.44 \\
54345.8140 & -0.65 & 0.38 \\
54346.7034 & 0.79 & 0.39 \\
54346.8011 & 0.47 & 0.56 \\
54390.6585 & 1.56 & 0.45 \\
54391.6343 & 4.24 & 0.55 \\
54392.6889 & 1.57 & 0.50 \\
54393.5114 & 2.16 & 0.38 \\
54393.7475 & 1.23 & 0.33 \\
54394.7116 & 1.96 & 0.41 \\
54421.5313 & -2.16 & 0.12 \\
54425.5706 & -1.17 & 0.42 \\
54447.5612 & -1.31 & 0.61 \\
54449.5587 & 0.18 & 0.52 \\
54451.5717 & 1.22 & 0.45 \\
54460.5745 & 3.44 & 0.57 \\
54461.6251 & 0.44 & 0.80 \\
54464.5623 & -0.52 & 0.46 \\
54660.9117 & -0.06 & 0.52 \\
54665.9245 & 0.23 & 0.48 \\
54672.9018 & 1.47 & 0.30 \\
54682.8605 & 3.75 & 0.50 \\
54701.8575 & -1.38 & 0.42 \\
54705.8143 & -2.51 & 0.72 \\
54775.6479 & 0.00 & 0.49 \\
54778.5665 & -0.08 & 0.41 \\
54825.5247 & -1.59 & 0.41 \\
54828.5473 & 0.36 & 0.80 \\
\hline \hline
\end{tabular}
\end{center}
\end{table}

\begin{table}
\caption{HARPS-TERRA velocity data of GJ 27.1.}\label{tab:gj27.1_terra}
\begin{center}
\begin{tabular}{lcccccc}
\hline \hline
Time & Velocity & Unc. \\

[JD-2400000] & [ms$^{-1}$] & [ms$^{-1}$] \\
\hline
54396.5474 & -8.35 & 1.45 \\
54397.5198 & -9.10 & 1.77 \\
54398.5162 & -5.05 & 1.31 \\
54398.8088 & -6.32 & 1.48 \\
54399.7018 & -5.59 & 1.26 \\
54400.5289 & 0.32 & 1.66 \\
54402.5416 & -2.14 & 1.78 \\
54403.5499 & -1.92 & 1.28 \\
54404.5258 & 3.04 & 3.54 \\
54404.6655 & -4.73 & 2.51 \\
54404.7956 & 0.07 & 3.56 \\
54420.5940 & 5.61 & 1.72 \\
54422.5626 & 1.85 & 1.60 \\
54424.5317 & -0.60 & 1.95 \\
54426.5621 & -5.83 & 1.40 \\
54428.6132 & -7.44 & 1.40 \\
54445.5804 & -6.31 & 1.62 \\
54447.6102 & -5.64 & 1.88 \\
54449.5706 & 0.00 & 2.44 \\
54450.5515 & 4.04 & 2.14 \\
54451.5887 & 8.07 & 1.59 \\
54453.5917 & 1.18 & 1.79 \\
54454.6166 & 2.67 & 1.22 \\
54456.5364 & -3.66 & 1.38 \\
54457.5401 & -7.06 & 1.34 \\
54458.5391 & -3.82 & 1.22 \\
54459.5818 & -4.02 & 1.64 \\
54460.6414 & 2.42 & 1.61 \\
54462.5392 & 1.96 & 2.62 \\
54463.5518 & 4.67 & 1.70 \\
54464.5853 & 1.27 & 1.62 \\
54479.5337 & 0.39 & 1.47 \\
54480.5346 & -0.06 & 1.43 \\
54481.5321 & 2.10 & 1.39 \\
54483.5328 & 7.79 & 1.93 \\
54484.5318 & 3.94 & 1.57 \\
54485.5315 & 3.84 & 2.05 \\
54486.5409 & 2.40 & 1.88 \\
54589.9044 & 3.93 & 2.02 \\
54590.9082 & 1.45 & 2.76 \\
54591.9063 & -4.55 & 2.46 \\
54592.9037 & -5.25 & 2.07 \\
54660.9233 & -0.61 & 2.50 \\
54661.9290 & -4.30 & 1.43 \\
54663.9120 & -6.41 & 1.71 \\
54665.9345 & -5.43 & 1.28 \\
54704.8504 & 4.64 & 1.67 \\
54707.8023 & -1.07 & 2.59 \\
54709.7946 & 2.83 & 2.14 \\
54833.5216 & 2.46 & 1.49 \\
\hline \hline
\end{tabular}
\end{center}
\end{table}

\begin{table}
\caption{HARPS-TERRA velocity data ofGJ 118.}\label{tab:gj118_terra}
\begin{center}
\begin{tabular}{lcc}
\hline \hline
Time & Velocity & Unc. \\

[JD-2400000] & [ms$^{-1}$] & [ms$^{-1}$] \\
\hline
54396.8106 & -0.06 & 1.15 \\
54397.6571 & -0.28 & 0.93 \\
54398.5893 & -6.37 & 1.55 \\
54399.8028 & -3.04 & 0.76 \\
54400.6373 & -2.39 & 1.12 \\
54400.8375 & -0.77 & 1.24 \\
54402.6226 & 0.00 & 1.17 \\
54402.8353 & 0.98 & 1.50 \\
54403.6109 & 2.07 & 1.00 \\
54403.7341 & 3.63 & 0.90 \\
54403.8602 & 3.53 & 1.15 \\
54404.6135 & 1.79 & 2.22 \\
54404.7191 & -0.62 & 1.42 \\
54404.8663 & 6.67 & 2.61 \\
\hline \hline
\end{tabular}
\end{center}
\end{table}

\begin{table}
\caption{HARPS-TERRA velocity data of GJ 160.2.}\label{tab:gj160.2_terra}
\begin{center}
\begin{tabular}{lcc}
\hline \hline
Time & Velocity & Unc. \\

[JD-2400000] & [ms$^{-1}$] & [ms$^{-1}$] \\
\hline
53007.5593 & -6.25 & 1.47 \\
53015.5965 & 2.18 & 1.38 \\
53360.6537 & 2.91 & 1.62 \\
53361.6926 & 2.95 & 1.48 \\
53729.7055 & -2.49 & 1.04 \\
54757.8682 & 1.04 & 1.25 \\
54902.4838 & -0.33 & 1.21 \\
\hline \hline
\end{tabular}
\end{center}
\end{table}

\begin{table}
\begin{center}
\caption{HARPS-TERRA velocity data of GJ 173.}\label{tab:gj173_terra}
\begin{tabular}{lcccccc}
\hline \hline
Time & Velocity & Unc. \\

[JD-2400000] & [ms$^{-1}$] & [ms$^{-1}$] \\
\hline
54396.8606 & 0.69 & 0.61 \\
54397.7662 & -0.11 & 0.59 \\
54398.7472 & -1.30 & 0.50 \\
54399.6894 & 0.00 & 0.62 \\
54400.7731 & 0.64 & 1.08 \\
\hline \hline
\end{tabular}
\end{center}
\end{table}

\begin{table}
\caption{HARPS-TERRA velocity data of GJ 180.}\label{tab:gj180_terra}
\begin{center}
\begin{tabular}{lcccccc}
\hline \hline
Time & Velocity & Unc. \\

[JD-2400000] & [ms$^{-1}$] & [ms$^{-1}$] \\
\hline
54455.5515 & -6.58 & 0.88 \\
54456.6814 & -5.59 & 0.78 \\
54457.5968 & -2.85 & 0.72 \\
54458.6863 & -4.35 & 0.82 \\
54460.6887 & 0.89 & 1.02 \\
54462.6088 & 1.45 & 0.94 \\
54463.5701 & 5.03 & 1.06 \\
54463.7659 & 6.30 & 1.53 \\
54464.6470 & 6.79 & 1.36 \\
54646.9336 & -6.12 & 1.27 \\
54661.9473 & -0.39 & 1.36 \\
54709.8978 & 0.00 & 1.06 \\
54710.8720 & 1.44 & 1.29 \\
54733.8431 & -2.84 & 0.92 \\
54734.8661 & 1.79 & 1.34 \\
54737.8605 & -0.05 & 1.55 \\
54750.8457 & -5.07 & 0.83 \\
54753.8227 & -0.71 & 0.97 \\
54758.7670 & 7.16 & 0.80 \\
54760.7606 & 1.32 & 0.84 \\
54762.7376 & 0.01 & 0.80 \\
54765.7222 & -2.56 & 1.85 \\
54768.6774 & -2.43 & 0.97 \\
54770.7836 & 2.32 & 0.86 \\
54775.7772 & 0.73 & 1.02 \\
54777.7177 & 0.01 & 0.94 \\
54778.6712 & 1.12 & 0.89 \\
54779.6771 & -2.89 & 1.16 \\
54800.7796 & -5.09 & 0.78 \\
54803.7127 & -0.33 & 1.07 \\
54806.6693 & 3.01 & 0.89 \\
\hline \hline
\end{tabular}
\end{center}
\end{table}

\begin{table}
\begin{center}
\caption{HARPS-TERRA velocity data of GJ 218.}\label{tab:gj218_terra}
\begin{tabular}{lcccccc}
\hline \hline
Time & Velocity & Unc. \\

[JD-2400000] & [ms$^{-1}$] & [ms$^{-1}$] \\
\hline
54455.6774 & 1.38 & 1.02 \\
54456.5799 & -0.89 & 0.99 \\
54457.6336 & 0.72 & 0.68 \\
54458.6459 & -0.62 & 0.69 \\
54461.6969 & -1.88 & 2.35 \\
54462.6559 & -1.03 & 0.84 \\
54462.8730 & 1.19 & 1.49 \\
54463.6353 & 1.01 & 0.85 \\
54464.6953 & 0.11 & 0.93 \\
\hline \hline
\end{tabular}
\end{center}
\end{table}

\begin{table}
\begin{center}
\caption{HARPS-TERRA velocity data of GJ 229.}\label{tab:gj229_terra}
\begin{tabular}{lcccccc}
\hline \hline
Time & Velocity & Unc. \\

[JD-2400000] & [ms$^{-1}$] & [ms$^{-1}$] \\
\hline
52986.7541 & -0.20 & 0.40 \\
53341.8225 & -0.51 & 0.44 \\
53366.7422 & -1.09 & 0.30 \\
53370.7396 & 0.04 & 0.37 \\
53372.6970 & 0.29 & 0.34 \\
53373.7291 & 0.00 & 0.24 \\
53375.7434 & -0.45 & 0.28 \\
53376.6843 & 0.52 & 0.29 \\
53377.6784 & 0.61 & 0.30 \\
53719.7740 & -4.42 & 0.72 \\
53719.7889 & -3.35 & 0.65 \\
53817.5349 & 0.32 & 0.51 \\
54172.5190 & -4.45 & 0.33 \\
54347.8764 & 1.94 & 0.34 \\
54464.7369 & -0.74 & 0.54 \\
54709.9085 & 0.05 & 0.48 \\
54710.9068 & 0.63 & 0.42 \\
\hline \hline
\end{tabular}
\end{center}
\end{table}

\begin{table}
\begin{center}
\caption{HARPS-TERRA velocity data of GJ 357.}\label{tab:gj357_terra}
\begin{tabular}{lcccccc}
\hline \hline
Time & Velocity & Unc. \\

[JD-2400000] & [ms$^{-1}$] & [ms$^{-1}$] \\
\hline
52986.8282 & 0.00 & 1.03 \\
53374.8370 & 2.48 & 0.51 \\
53517.4976 & -1.09 & 0.75 \\
53518.4946 & -0.43 & 0.67 \\
53815.7227 & 2.96 & 0.44 \\  
\hline \hline
\end{tabular}
\end{center}
\end{table}

\begin{table}
\begin{center}
\caption{HARPS-TERRA velocity data of GJ 377.}\label{tab:gj377_terra}
\begin{tabular}{lcccccc}
\hline \hline
Time & Velocity & Unc. \\

[JD-2400000] & [ms$^{-1}$] & [ms$^{-1}$] \\
\hline
54455.8469 & 18.10 & 5.97 \\
54456.7377 & 11.15 & 4.26 \\
54457.7906 & 15.91 & 4.61 \\
54458.7920 & 13.05 & 4.49 \\
54461.8412 & 3.55 & 6.12 \\
54462.7193 & 11.49 & 4.56 \\
54464.8349 & 7.35 & 4.80 \\
54658.4735 & 0.00 & 5.91 \\
\hline \hline
\end{tabular}
\end{center}
\end{table}

\begin{table}
\begin{center}
\caption{HARPS-TERRA velocity data of GJ 422.}\label{tab:gj422_terra}
\begin{tabular}{lcccccc}
\hline \hline
Time & Velocity & Unc. \\

[JD-2400000] & [ms$^{-1}$] & [ms$^{-1}$] \\
\hline
53033.8034 & -7.14 & 1.09 \\
53049.8003 & 2.86 & 1.36 \\
53145.6175 & 0.00 & 2.28 \\
53153.5341 & 2.49 & 1.75 \\
53406.8387 & -6.87 & 0.96 \\
53781.8021 & 3.10 & 0.86 \\
54143.7926 & 4.55 & 0.95 \\
54488.8224 & -1.19 & 0.96 \\
54541.7367 & -0.77 & 0.87 \\
54547.7139 & -0.93 & 0.92 \\
54565.6547 & 4.25 & 1.34 \\
54582.6565 & -3.73 & 0.96 \\
54583.4919 & 0.67 & 2.16 \\
54588.5603 & 8.23 & 1.34 \\
54589.5131 & 8.16 & 1.10 \\
54590.5245 & 7.71 & 1.23 \\
54591.5064 & 5.69 & 1.13 \\
54592.5170 & 3.47 & 1.17 \\
54658.5135 & -6.98 & 1.87 \\
54660.5477 & -2.97 & 2.87 \\
54662.4624 & 0.83 & 1.29 \\
54665.4896 & -0.89 & 1.17 \\
54873.7804 & -6.39 & 0.95 \\
55310.6933 & -4.38 & 5.29 \\
55315.6795 & -3.13 & 1.75 \\
\hline \hline
\end{tabular}
\end{center}
\end{table}

\begin{table}
\begin{center}
\caption{HARPS-TERRA velocity data of GJ 433.}\label{tab:gj433_terra}
\begin{tabular}{lcccccc}
\hline \hline
Time & Velocity & Unc. \\

[JD-2400000] & [ms$^{-1}$] & [ms$^{-1}$] \\
\hline
52989.8351 & 0.99 & 1.03 \\
52996.8440 & -1.23 & 0.37 \\
53511.5748 & -3.89 & 0.74 \\
53516.5922 & -3.83 & 1.19 \\
53516.5970 & -5.84 & 1.49 \\
53520.6016 & 8.13 & 2.44 \\
53809.7514 & -1.20 & 0.58 \\
53810.7295 & -5.55 & 0.50 \\
53817.7708 & -2.15 & 0.71 \\
54134.8273 & -1.21 & 0.63 \\
54200.6757 & -0.09 & 0.64 \\
54229.6428 & 2.32 & 0.95 \\
54256.5545 & 5.58 & 0.78 \\
54257.5314 & 7.06 & 0.76 \\
54258.4960 & 2.87 & 0.73 \\
54296.5634 & 2.64 & 1.07 \\
54299.5360 & 3.00 & 1.41 \\
54459.8498 & -3.93 & 0.62 \\
54460.8517 & -3.92 & 0.92 \\
54526.7455 & -3.31 & 0.56 \\
54549.6733 & 4.67 & 1.00 \\
54552.6940 & 4.47 & 0.53 \\
54556.6397 & 0.43 & 0.72 \\
54562.6851 & -2.43 & 0.58 \\
54566.6494 & 3.83 & 0.62 \\
54570.6111 & -2.75 & 0.65 \\
54639.5479 & 0.00 & 0.86 \\
54640.5364 & 3.61 & 0.72 \\
54641.5082 & 2.70 & 0.67 \\
54642.5316 & 1.07 & 0.73 \\
54643.5411 & -0.53 & 0.71 \\
54645.5102 & -0.04 & 0.96 \\
54646.5103 & 4.17 & 0.93 \\
54647.4706 & 6.91 & 0.49 \\
54648.5072 & 7.77 & 0.66 \\
54658.4623 & -1.03 & 0.63 \\
54660.4601 & -0.12 & 1.03 \\
54661.4629 & 2.30 & 0.74 \\
54662.4738 & 3.43 & 0.72 \\
54663.4617 & 2.04 & 0.61 \\
54664.4669 & 2.25 & 1.21 \\
54665.4710 & -1.43 & 0.63 \\
54666.4653 & -0.77 & 0.66 \\
54672.5149 & -1.71 & 1.08 \\
54674.4708 & -5.46 & 1.02 \\
54677.4775 & 1.63 & 0.87 \\
54678.4730 & 2.86 & 2.02 \\
54679.4762 & 0.58 & 0.61 \\
54681.4690 & 0.45 & 1.03 \\
54682.4744 & 0.66 & 0.59 \\
55041.4917 & -1.12 & 0.66 \\
55046.4638 & 5.42 & 0.88 \\
55047.4827 & 5.45 & 1.05 \\
55048.4806 & -2.12 & 0.63 \\
55049.4866 & -3.00 & 1.15 \\
55050.4783 & -3.40 & 2.95 \\
55053.4761 & -1.96 & 0.82 \\
55054.4811 & 2.68 & 1.76 \\
55055.4653 & -0.80 & 0.98 \\
55056.4586 & -3.60 & 0.93 \\
55057.4750 & -3.82 & 1.15 \\
55234.7242 & -1.70 & 0.78 \\
\hline \hline
\end{tabular}
\end{center}
\end{table}

\begin{table}
\begin{center}
\caption{HARPS-TERRA velocity data of GJ 510.}\label{tab:gj510_terra}
\begin{tabular}{lcccccc}
\hline \hline
Time & Velocity & Unc. \\

[JD-2400000] & [ms$^{-1}$] & [ms$^{-1}$] \\
\hline
54565.7787 & 4.22 & 1.82 \\
54583.6781 & 3.10 & 0.91 \\
54587.7166 & 0.00 & 2.49 \\
54588.6969 & -0.70 & 1.37 \\
54589.6525 & 0.01 & 1.31 \\
54590.6470 & -1.59 & 1.19 \\
54591.6444 & -3.39 & 1.13 \\
54592.6910 & -2.39 & 0.94 \\
54660.5126 & 2.34 & 3.04 \\
\hline \hline
\end{tabular}
\end{center}
\end{table}

\begin{table}
\begin{center}
\caption{HARPS-TERRA velocity data of 551.}\label{tab:proxima_terra}
\begin{tabular}{lcccccc}
\hline \hline
Time & Velocity & Unc. \\

[JD-2400000] & [ms$^{-1}$] & [ms$^{-1}$] \\
\hline
53152.5998 & -1.14 & 0.61 \\
53202.5848 & -4.07 & 3.04 \\
53203.5372 & -1.19 & 0.95 \\
53205.5317 & -0.10 & 3.78 \\
53207.5052 & 0.23 & 3.40 \\
53207.5109 & 1.46 & 3.67 \\
53577.4956 & 2.57 & 0.69 \\
53809.8936 & -2.69 & 0.66 \\
53810.7835 & -2.51 & 0.52 \\
53812.7822 & 0.11 & 0.76 \\
53816.8096 & -4.49 & 0.57 \\
54173.8182 & -0.76 & 0.67 \\
54293.5865 & 2.46 & 0.66 \\
54296.5934 & -0.18 & 0.74 \\
54299.5978 & 5.33 & 1.04 \\
54300.5576 & -1.97 & 0.58 \\
54646.5693 & 0.99 & 0.80 \\
54658.5644 & 1.71 & 1.00 \\
54666.5435 & 0.00 & 0.60 \\
54878.8638 & -0.07 & 0.55 \\
54879.8301 & 0.70 & 0.76 \\
54881.8737 & -0.76 & 0.52 \\
54882.8535 & -1.48 & 0.64 \\
54883.8616 & 0.19 & 0.90 \\
54884.8354 & 1.34 & 0.70 \\
54885.8751 & 0.62 & 0.62 \\
54886.8436 & 0.28 & 0.68 \\
\hline \hline
\end{tabular}
\end{center}
\end{table}

\begin{table}
\begin{center}
\caption{HARPS-TERRA velocity data of GJ 620.}\label{tab:gj620_terra}
\begin{tabular}{lcccccc}
\hline \hline
Time & Velocity & Unc. \\

[JD-2400000] & [ms$^{-1}$] & [ms$^{-1}$] \\
\hline
54565.9166 & 3.21 & 1.29 \\
54660.6863 & 3.66 & 1.36 \\
54661.5910 & 3.03 & 0.95 \\
54662.5743 & -0.04 & 1.30 \\
54663.5939 & -3.78 & 0.72 \\
54664.5941 & -2.26 & 1.05 \\
54665.5853 & 0.00 & 0.85 \\
54672.5754 & -0.03 & 1.03 \\
54674.5383 & 7.54 & 1.10 \\
54679.5310 & 0.53 & 1.33 \\
54681.5307 & -0.54 & 1.61 \\
\hline \hline
\end{tabular}
\end{center}
\end{table}

\begin{table}
\begin{center}
\caption{HARPS-TERRA velocity data of GJ 637.}\label{tab:gj637_terra}
\begin{tabular}{lcccccc}
\hline \hline
Time & Velocity & Unc. \\

[JD-2400000] & [ms$^{-1}$] & [ms$^{-1}$] \\
\hline
54566.8654 & 1.64 & 3.23 \\
54658.7124 & 2.86 & 2.31 \\
54660.6238 & 1.00 & 2.71 \\
54661.7459 & -0.45 & 1.98 \\
54662.6506 & -2.75 & 1.82 \\
54664.7525 & -1.08 & 2.98 \\
54665.7604 & -2.78 & 1.86 \\
54666.6692 & 1.56 & 2.10 \\
\hline \hline
\end{tabular}
\end{center}
\end{table}

\begin{table}
\begin{center}
\caption{HARPS-TERRA velocity data of GJ 682.}\label{tab:gj682_terra}
\begin{tabular}{lcccccc}
\hline \hline
Time & Velocity & Unc. \\

[JD-2400000] & [ms$^{-1}$] & [ms$^{-1}$] \\
\hline
53158.7761 & 0.12 & 2.72 \\
53205.6737 & -3.93 & 1.19 \\
53484.8539 & 1.10 & 0.65 \\
53489.8754 & 3.97 & 1.17 \\
53511.8240 & -1.14 & 1.12 \\
53814.8765 & -1.95 & 0.47 \\
53815.8679 & -1.23 & 0.48 \\
53974.5431 & -4.01 & 0.92 \\
54257.7558 & 0.00 & 0.59 \\
54258.7181 & -0.93 & 0.51 \\
54658.6872 & 0.34 & 0.78 \\
54666.7448 & 2.85 & 0.64 \\
\hline \hline
\end{tabular}
\end{center}
\end{table}

\begin{table}
\caption{HARPS-TERRA velocity data of GJ 699}\label{tab:barnard_terra}
\begin{center}
\begin{tabular}{lcc}
\hline \hline
Time & Velocity & Unc. \\

[JD-2400000] & [ms$^{-1}$] & [ms$^{-1}$] \\
\hline
54194.8939 & -0.48 & 0.38 \\
54196.8834 & 0.46 & 0.43 \\
54197.8948 & -1.01 & 0.44 \\
54198.8903 & 1.29 & 0.51 \\
54199.9182 & 0.00 & 0.37 \\
54200.9206 & -0.83 & 0.35 \\
54202.9187 & 0.52 & 0.38 \\
54227.8438 & 0.80 & 0.44 \\
54291.5969 & -1.46 & 0.43 \\
54298.7145 & 0.07 & 0.64 \\
54315.6613 & 2.28 & 0.73 \\
54316.6473 & 0.71 & 0.57 \\
54319.6380 & 0.35 & 0.45 \\
54320.6502 & -0.71 & 0.39 \\
54340.6506 & -0.80 & 0.34 \\
54343.5572 & -0.84 & 0.32 \\
54347.5581 & -1.12 & 0.39 \\
54385.5239 & 1.31 & 0.86 \\
54389.5016 & 2.26 & 1.23 \\
54523.8813 & -1.24 & 0.59 \\
54551.9090 & -1.56 & 0.48 \\
54588.9210 & -1.86 & 0.59 \\
\hline \hline
\end{tabular}
\end{center}
\end{table}

\clearpage

\begin{table}
\begin{center}
\caption{HARPS-TERRA velocity data of GJ 739.}\label{tab:gj739_terra}
\begin{tabular}{lcccccc}
\hline \hline
Time & Velocity & Unc. \\

[JD-2400000] & [ms$^{-1}$] & [ms$^{-1}$] \\
\hline
54658.7790 & 0.75 & 1.29 \\
54660.8304 & -0.75 & 1.20 \\
\hline \hline
\end{tabular}
\end{center}
\end{table}

\clearpage

\begin{table}
\begin{center}
\caption{HARPS-TERRA velocity data of GJ 821.}\label{tab:gj821_terra}
\begin{tabular}{lcccccc}
\hline \hline
Time & Velocity & Unc. \\

[JD-2400000] & [ms$^{-1}$] & [ms$^{-1}$] \\
\hline
54641.9059 & 0.06 & 1.38 \\
54661.8214 & -1.02 & 0.84 \\
54662.8109 & 2.01 & 1.27 \\
54663.8384 & -0.58 & 1.13 \\
54664.8407 & -0.47 & 1.41 \\    
\hline \hline
\end{tabular}
\end{center}
\end{table}

\begin{table}
\begin{center}
\caption{HARPS-TERRA velocity data of GJ 842.}\label{tab:gj842_terra}
\begin{tabular}{lcccccc}
\hline \hline
Time & Velocity & Unc. \\

[JD-2400000] & [ms$^{-1}$] & [ms$^{-1}$] \\
\hline
54646.9186 & 0.73 & 1.08 \\
54660.9503 & 2.00 & 0.85 \\
54661.9071 & -0.94 & 0.70 \\
54662.8917 & -0.99 & 0.71 \\
54663.8893 & 0.69 & 0.59 \\
54664.8921 & -0.80 & 1.04 \\
54665.8894 & -0.67 & 0.72 \\
\hline \hline
\end{tabular}
\end{center}
\end{table}

\begin{table}
\begin{center}
\caption{HARPS-TERRA velocity data of GJ 855.}\label{tab:gj855_terra}
\begin{tabular}{lcccccc}
\hline \hline
Time & Velocity & Unc. \\

[JD-2400000] & [ms$^{-1}$] & [ms$^{-1}$] \\
\hline
54312.8464 & -1.10 & 0.75 \\
54640.9184 & 5.96 & 1.77 \\
54660.9015 & -5.95 & 1.17 \\
54661.9188 & -6.94 & 1.09 \\
54662.9035 & -0.81 & 0.89 \\
54663.9017 & 0.00 & 1.14 \\
54664.9034 & 7.56 & 1.50 \\
54665.9016 & 3.95 & 1.22 \\
54805.5244 & -0.96 & 1.05 \\
55429.8448 & 4.17 & 1.24 \\
\hline \hline
\end{tabular}
\end{center}
\end{table}

\begin{table}
\begin{center}
\caption{HARPS-TERRA velocity data of GJ 911.}\label{tab:gj911_terra}
\begin{tabular}{lcccccc}
\hline \hline
Time & Velocity & Unc. \\

[JD-2400000] & [ms$^{-1}$] & [ms$^{-1}$] \\
\hline
53295.7331 & -0.92 & 1.43 \\
53307.5889 & 1.83 & 1.41 \\
54751.7125 & -0.91 & 1.82 \\
\hline \hline
\end{tabular}
\end{center}
\end{table}

\begin{table}
\begin{center}
\caption{HARPS-TERRA velocity data of GJ 1009.}\label{tab:gj1009_terra}
\begin{tabular}{lcccccc}
\hline \hline
Time & Velocity & Unc. \\

[JD-2400000] & [ms$^{-1}$] & [ms$^{-1}$] \\
\hline
54396.5242 & -3.28 & 1.05 \\
54397.5087 & -5.28 & 1.01 \\
54398.5041 & -4.05 & 0.86 \\
54399.5048 & -1.20 & 0.89 \\
54400.5178 & -1.29 & 1.15 \\
54402.5541 & 2.50 & 1.23 \\
54402.7534 & 2.95 & 1.07 \\
54403.5622 & 0.00 & 0.78 \\
54403.7606 & -1.91 & 1.08 \\
54404.5388 & 0.75 & 2.20 \\
54404.6831 & 1.04 & 1.30 \\
54404.7599 & 2.87 & 2.53 \\
\hline \hline
\end{tabular}
\end{center}
\end{table}

\begin{table}
\begin{center}
\caption{HARPS-TERRA velocity data of GJ 1100.}\label{tab:gj1100_terra}
\begin{tabular}{lcccccc}
\hline \hline
Time & Velocity & Unc. \\

[JD-2400000] & [ms$^{-1}$] & [ms$^{-1}$] \\
\hline
54455.7488 & -0.53 & 3.32 \\
54456.7074 & 0.00 & 1.85 \\
54457.7099 & 0.31 & 1.09 \\
54458.8622 & -1.70 & 2.08 \\
54459.6988 & 0.53 & 1.11 \\
54460.7429 & 2.00 & 1.72 \\
54462.7733 & 0.08 & 1.39 \\
54463.6757 & -1.88 & 1.51 \\
54463.8467 & -0.74 & 1.74 \\
54464.7948 & -0.84 & 1.53 \\
\hline \hline
\end{tabular}
\end{center}
\end{table}

\begin{table}
\begin{center}
\caption{HARPS-TERRA velocity data of GJ 3671.}\label{tab:gj3671_terra}
\begin{tabular}{lcccccc}
\hline \hline
Time & Velocity & Unc. \\

[JD-2400000] & [ms$^{-1}$] & [ms$^{-1}$] \\
\hline
54565.7160 & -0.39 & 1.49 \\
54583.5183 & 0.39 & 1.40 \\
\hline \hline
\end{tabular}
\end{center}
\end{table}

\begin{table}
\begin{center}
\caption{HARPS-TERRA velocity data of GJ 3759.}\label{tab:gj3759_terra}
\begin{tabular}{lcccccc}
\hline \hline
Time & Velocity & Unc. \\

[JD-2400000] & [ms$^{-1}$] & [ms$^{-1}$] \\
\hline
54566.7038 & 1.07 & 1.88 \\
54583.6379 & 0.06 & 1.13 \\
54586.8407 & 6.78 & 4.02 \\
54586.8526 & -2.44 & 3.33 \\
54588.6701 & -5.45 & 1.25 \\
54589.6239 & 0.60 & 1.34 \\
54590.6215 & -1.65 & 1.08 \\
54591.6178 & -2.35 & 1.29 \\
54592.6502 & -1.60 & 1.01 \\
54658.5268 & 4.26 & 1.35 \\
54666.5303 & 0.00 & 0.83 \\
\hline \hline
\end{tabular}
\end{center}
\end{table}

\begin{table}
\begin{center}
\caption{HARPS-TERRA velocity data of GJ 3973.}\label{tab:gj3973_terra}
\begin{tabular}{lcccccc}
\hline \hline
Time & Velocity & Unc. \\

[JD-2400000] & [ms$^{-1}$] & [ms$^{-1}$] \\
\hline
54570.8523 & 1.39 & 2.06 \\
54661.6505 & -0.44 & 1.31 \\
54662.7362 & 1.13 & 1.37 \\
54663.6587 & -0.73 & 0.98 \\
54664.6193 & -1.02 & 1.34 \\
54665.6492 & 0.55 & 0.93 \\
54666.6322 & -0.88 & 0.93 \\
\hline \hline
\end{tabular}
\end{center}
\end{table}

\label{lastpage}


\begin{thebibliography}{100}
\bibitem[\protect\astroncite{Anglada-Escud\'e et al.}{2012}]{anglada2012} Anglada-Escud\'e, G., Arriagada, P., Vogt, S. S., et al. 2012, ApJ, 751, L16
\bibitem[\protect\astroncite{Anglada-Escud\'e et al.}{2013}]{anglada2013} Anglada-Escud\'e, G., Tuomi, M., Gerlach, E., et al. 2013, A\&A, 556, A126
\bibitem[\protect\astroncite{Anglada-Escud\'e \& Butler}{2012}]{anglada2012c} Anglada-Escud\'e, G. \& Butler, R. P. 2012, ApJS, 200, 15
\bibitem[\protect\astroncite{Anglada-Escud\'e \& Tuomi}{2012}]{anglada2012b} Anglada-Escud\'e, G. \& Tuomi, M. 2012, A\&A, 548, A58
\bibitem[\protect\astroncite{Avenhaus et al.}{2012}]{avenhaus2012} Avenhaus, H., Schmid, H. M., \& Meyer, M. R. 2012, A\&A, 548, A105
\bibitem[\protect\astroncite{Baluev}{2009}]{baluev2009} Baluev, R. V., 2012, MNRAS, 393, 969
\bibitem[\protect\astroncite{Baluev}{2013}]{baluev2012} Baluev, R. V. 2013, MNRAS, 429, 2052
\bibitem[\protect\astroncite{Berger}{1980}]{berger1980} Berger, J. O, 1980, Statistical Decision Theory and Bayesian Analysis (Springer)
\bibitem[\protect\astroncite{Boisse et al.}{2011}]{boisse2011} Boisse, I., Bouchy, F., H\'ebrard, G., et al. 2011, A\&A, 528, A4
\bibitem[\protect\astroncite{Bonfils et al.}{2005}]{bonfils2005} Bonfils, X., Forveille, T., Delfosse, X., et al. 2005, A\&A, 443, L15
\bibitem[\protect\astroncite{Bonfils et al.}{2013a}]{bonfils2013} Bonfils, X., Delfosse, X., Udry, S., et al. 2013a, A\&A, 549, A109
\bibitem[\protect\astroncite{Bonfils et al.}{2013b}]{bonfils2013b} Bonfils, X., Lo Curto, G., Correia, A. C. M., et al. 2013b, A\&A, 556, A110
\bibitem[\protect\astroncite{Boyajian et al.}{2012}]{boyajian2012} Boyajian, T. S., von Braun, K., van Belle, G., et al. 2012, ApJ, 757, 112
\bibitem[\protect\astroncite{Casagrande et al.}{2008}]{casagrande2008} Casagrande, L., Flynn, C., \& Bessel, M. 2008, MNRAS, 389, 585
\bibitem[\protect\astroncite{Choi et al.}{2012}]{choi2012} Choi, J., McCarthy, C., Marcy, G. W. et al. 2013, ApJ, 764, 131
\bibitem[\protect\astroncite{Delfosse et al.}{2013}]{delfosse2012} Delfosse, X., Bonfils, X., Forveille, T., et al. 2013, A\&A, 553, A8
\bibitem[\protect\astroncite{Dieterich et al.}{2012}]{dieterich2012} Dieterich, S. B., Henry, T. J., Golimowski, D. A., et al. 2012, AJ, 144, 64
\bibitem[\protect\astroncite{Dressing \& Charbonneau}{2013}]{dressing2013} Dressing, C. D. \& Charbonneau, D. 2013, ApJ, 767, 95
\bibitem[\protect\astroncite{Endl et al.}{2006}]{endl2006} Endl, M., Cochran, W. D., K\"urster, M., et al. 2006, ApJ, 649, 436
\bibitem[\protect\astroncite{Endl \& K\"urster}{2008}]{endl2008} Endl, M. \& K\"urster, M. 2008, A\&A, 488, 1149
\bibitem[\protect\astroncite{Evett}{1991}]{evett1991} Evett, I. W. 1991, Implementing Bayesian methods in forensic science, Paper presented at the Fourth Valencia International Meeting on Bayesian Statistics
\bibitem[\protect\astroncite{Feroz et al.}{2011}]{feroz2011} Feroz, F., Balan, S. T., \& Hobson, M. P. 2011, MNRAS, 415, 3462
\bibitem[\protect\astroncite{Ford}{2005}]{ford2005} Ford, E. B. 2005, AJ, 129, 1706
\bibitem[\protect\astroncite{Ford}{2006}]{ford2006} Ford, E. B. 2006, ApJ, 642, 505
\bibitem[\protect\astroncite{Forveille et al.}{2011}]{forveille2011} Forveille, T., Bonfils, X., Lo Curto, G., et al. 2011, A\&A, 526, A141
\bibitem[\protect\astroncite{Fressing et al.}{2013}]{fressin2013} Fressin, F., Torres, G., Charbonneau, D., et al. 2013, ApJ, 766, 81
\bibitem[\protect\astroncite{Gelman \& Rubin}{1992}]{gelman1992} Gelman, A. \& Rubin, D. B. 1992, Statistical Science, 7, 457
\bibitem[\protect\astroncite{Gray et al.}{2006}]{gray2006} Gray, R. O., Corbally, C. J., Garrison, R. F., et al. 2006, AJ, 132, 161
\bibitem[\protect\astroncite{Gregory}{2011}]{gregory2011a} Gregory, P. C. 2011a, MNRAS, 410, 94
\bibitem[\protect\astroncite{Gregory}{2011}]{gregory2011} Gregory, P. C. 2011b, MNRAS, 415, 2523
\bibitem[\protect\astroncite{Haario et al.}{2001}]{haario2001} Haario, H., Saksman, E., \& Tamminen, J. 2001, Bernoulli, 7, 223
\bibitem[\protect\astroncite{Hastings}{1970}]{hastings1970} Hastings, W. 1970, Biometrika 57, 97
\bibitem[\protect\astroncite{Houdebine}{2011}]{houdebine2011} Houdebine, E. R. 2011, MNRAS, 416, 2233
\bibitem[\protect\astroncite{Howard et al.}{2012}]{howard2012} Howard, A. W., Marcy, G. W., Bryson, S. T., et al. 2012, ApJS, 201, 15
\bibitem[\protect\astroncite{Jeffreys}{1961}]{jeffreys1961} Jeffreys, H. 1961, The Theory of Probability (The Oxford University Press)
\bibitem[\protect\astroncite{Jenkins et al.}{2009}]{jenkins2009} Jenkins, J. S., Ramsey, L. W., Jones, H. R. A., et al. 2009, ApJ, 704, 975
\bibitem[\protect\astroncite{Jenkins et al.}{2013a}]{jenkins2013a} Jenkins, J. S., Tuomi, M., Brasser, R., et al. 2013a, ApJ, 771, 41
\bibitem[\protect\astroncite{Jenkins et al.}{2013b}]{jenkins2013b} Jenkins, J. S., Tuomi, M., Jones, H. R. A., et al. 2013b, in preparation
\bibitem[\protect\astroncite{Kass \& Raftery}{1995}]{kass1995} Kass, R. E. \& Raftery, A. E. 1995, J. Am. Stat. Ass., 430, 773
\bibitem[\protect\astroncite{Kaipio \& Somersalo}{2005}]{kaipio2005} Kaipio, J. \& Somersalo E. 2005, Statistical and Computational Inverse Problems, Applied Mathematical Sciences 160
\bibitem[\protect\astroncite{Kipping}{2013}]{kipping2013} Kipping, D. M. 2013, MNRAS, 434, L51
\bibitem[\protect\astroncite{Koen et al.}{2010}]{koen2010} Koen, C., Kilkenny, D., van Wyk, F., \& Marang, F. 2010, MNRAS, 403, 1949
\bibitem[\protect\astroncite{Kopparapu et al.}{2013}]{kopparapu2013} Kopparapu, R. K., Ramirez, R., Kasting, J. F., et al. 2013, ApJ, 765, 131
\bibitem[\protect\astroncite{Kopparapu}{2013}]{kopparapu2013b} Kopparapu, R. K. 2013, ApJ, 767, L8
\bibitem[\protect\astroncite{K\"urster et al.}{2008}]{kurster2008} K\"urster, M., Endl, M., \& Reffert, S. 2008, A\&A, 483, 869
\bibitem[\protect\astroncite{Lestrade et al.}{2009}]{lestrade2009} Lestrade, J.-F., Wyatt, M. C., Bertoldi, F., et al. 2009, A\&A, 506, 1455
\bibitem[\protect\astroncite{Lomb}{1976}]{lomb1976} Lomb, N. R. 1976, Astrophys. Space Sci., 39, 447
\bibitem[\protect\astroncite{Loredo et al.}{2012}]{loredo2012} Loredo, T. J., Berger, J. O., Chernoff, D. F., et al. 2012, Stat. Met., 9, 101
\bibitem[\protect\astroncite{Lopez \& Jenkins}{2012}]{lopez2012} Lopez, S. \& Jenkins, J. S. 2012, ApJ, 756, 177
\bibitem[\protect\astroncite{Mayor et al.}{2009}]{mayor2009} Mayor, M., Bonfils, X., Forveille, T., et al. 2009, A\&A, 507, 487
\bibitem[\protect\astroncite{Metropolis et al.}{1953}]{metropolis1953} Metropolis, N., Rosenbluth, A. W., Rosenbluth, M. N., et al. 1953, J. Chem. Phys., 21, 1087
\bibitem[\protect\astroncite{Montes et al.}{2001}]{montes2001} Montes, D., L\'opez-Santiago, J., G\'alvez, M. C., et al. 2001, MNRAS, 328, 45
\bibitem[\protect\astroncite{Montet et al.}{2014}]{montet2013} Montet, B. T., Crepp, J. R., Johnson, J. A., et al. 2014, ApJ, 781, 28
\bibitem[\protect\astroncite{Morton \& Johnson}{2011}]{morton2011} Morton, T. D. \& Johnson J. A. 2011, ApJ, 738, 170
\bibitem[\protect\astroncite{Morton \& Swift}{2013}]{morton2013} Morton, T. D. \& Swift, J. 2013, submitted to ApJ, (arXiv:1303.3013)
\bibitem[\protect\astroncite{Nakajima et al.}{1995}]{nakajima1995} Nakajima, T., Oppenheimer, B. R., Kulkarni, S. R., et al. 1995, Nature, 378, 463
\bibitem[\protect\astroncite{Newton \& Raftery}{1994}]{newton1994} Newton, M. A. \& Raftery, A. E. 1994, JRSS B, 56, 3
\bibitem[\protect\astroncite{Queloz et al.}{2001}]{queloz2001} Queloz, D., Henry, G. W., Sivan, J. P., et al. 2001, A\&A, 379, 279
\bibitem[\protect\astroncite{Reid et al.}{2004}]{reid2004} Reid, I. N., Cruz, K. L., Allen, P., et al. 2004, AJ, 128, 463
\bibitem[\protect\astroncite{Reiners}{2007}]{reiners2007} Reiners, A 2007, A\&A, 467, 259
\bibitem[\protect\astroncite{Rivera et al.}{2010}]{rivera2010} Rivera, E. J., Laughlin, G., Butler, R. P., et al. 2010, ApJ, 719, 890
\bibitem[\protect\astroncite{Scargle}{1982}]{scargle1982} Scargle, J. D. 1982, ApJ, 263, 835
\bibitem[\protect\astroncite{Schneider et al.}{2011}]{schneider2011} Schneider, J., Dedieu, C., Le Sinader, P., et al. 2011, A\&A, 532, A79
\bibitem[\protect\astroncite{Selsis et al.}{2007}]{selsis2007} Selsis, F., Kasting, J. F., Levrard, B., et al. 2007, A\&A, 476, 1373
\bibitem[\protect\astroncite{Torres et al.}{2006}]{torres2006} Torres, C. A. O., Quast, G. R., Da Silva, L., et al. 2006, A\&A, 460, 695
\bibitem[\protect\astroncite{Trotta}{2007}]{trotta2007} Trotta, R. 2007, MNRAS, 378, 72
\bibitem[\protect\astroncite{Tuomi}{2011}]{tuomi2011} Tuomi, M. 2011, A\&A, 528, L5
\bibitem[\protect\astroncite{Tuomi}{2012}]{tuomi2012} Tuomi, M. 2012, A\&A, 543, A52
\bibitem[\protect\astroncite{Tuomi \& Anglada-Escud\'e}{2013}]{tuomi2013c} Tuomi, M. \& Anglada-Escud\'e 2013, A\&A, 556, A111
\bibitem[\protect\astroncite{Tuomi \& Jenkins}{2012}]{tuomi2012c} Tuomi, M. \& Jenkins, J. S. 2012, submitted to A\&A (arXiv:1211.1280)
\bibitem[\protect\astroncite{Tuomi et al.}{2013a}]{tuomi2013a} Tuomi, M., Anglada-Escud\'e, G., Gerlach, E., et al. 2013a A\&A, 549, A48
\bibitem[\protect\astroncite{Tuomi et al.}{2013b}]{tuomi2013b} Tuomi, M., Jones, H. R. A., Jenkins, J. S., et al. 2013b, A\&A, 551, A79
\bibitem[\protect\astroncite{Udry et al.}{2007}]{udry2007} Udry, S., Bonfils, X., Delfosse, X., et al. 2007, A\&A, 469, L43
\bibitem[\protect\astroncite{van Leeuwen}{2007}]{vanleeuwen2007} van Leeuwen, F., 2007, A\&A, 474, 653
\bibitem[\protect\astroncite{Vogt et al.}{2010}]{vogt2010} Vogt, S., Butler, P., Rivera, E., et al. 2010. ApJ, 723, 954.
\bibitem[\protect\astroncite{Vogt et al.}{2012}]{vogt2012} Vogt, S. S., Butler, R. P., \& Haghighipour, R. N. 2012, AN, 333, 1
\bibitem[\protect\astroncite{Walkowicz \& Hawley}{2009}]{walkowicz2009} Walkowicz, L. M. \& Hawley, S. L. 2009, AJ, 137, 3297
\bibitem[\protect\astroncite{Zakamska et al.}{2011}]{zakamska2011} Zakamska, N. L., Pan, M., \& Ford, E. B. 2011, MNRAS, 410, 1895
\bibitem[\protect\astroncite{Zechmeister et al.}{2009}]{zechmeister2009} Zechmeister, M., K\"urster, M., \& Endl, M. 2009, A\&A, 505, 859
\end{thebibliography}
\end{document}